\documentclass[12pt]{iopart}
\usepackage{dcolumn,graphicx,color,booktabs,microtype,afterpage}
\usepackage{float}
\usepackage{amssymb}
\usepackage{url}  % To allow also url-s with underscore
\usepackage{sidecap}
\usepackage[mathlines]{lineno}
\usepackage{citesort}
\begin{document}

%\title[Enhanced $T_c$ and preserved TRS in the fully-gapped ReBe$_{22}$ superconductor]{Enhanced $T_c$ and preserved time-reversal symmetry in the fully-gapped ReBe$_{22}$ superconductor}
\title[Enhanced $T_c$ and multiband SC in the fully-gapped ReBe$_{22}$ superconductor]{Enhanced $T_c$ and multiband superconductivity in the fully-gapped ReBe$_{22}$ superconductor}

\author{T.\ Shang$^1$, A.\ Amon$^2$, D.\ Kasinathan$^2$, W.\ Xie$^3$, M.\ Bobnar$^2$, Y.~ Chen$^3$, A.\ Wang$^3$, M.\ Shi$^4$, M.\ Medarde$^1$, H.\ Q.\ Yuan$^{3,5}$, T.~Shiroka$^{6,7}$}

\address{$^1$Laboratory for Multiscale Materials Experiments, Paul Scherrer Institut, Villigen CH-5232, Switzerland}
\address{$^2$Max-Planck-Institut f\"ur Chemische Physik fester Stoffe, Dresden, 01187, Germany}
\address{$^3$Center for Correlated Matter and Department of Physics, Zhejiang University, Hangzhou, 310058, China}
\address{$^4$Swiss Light Source, Paul Scherrer Institut, Villigen CH-5232, Switzerland}
\address{$^5$Collaborative Innovation Center of Advanced Microstructures, Nanjing, 210093, China}
\address{$^6$Laboratorium f\"ur Festk\"orperphysik, ETH Z\"urich, CH-8093 Zurich, Switzerland}
\address{$^7$Laboratory for Muon-Spin Spectroscopy, Paul Scherrer Institut, CH-5232 Villigen PSI, Switzerland}
\ead{tian.shang@psi.ch and tshiroka@phys.ethz.ch}
%\ead{tshiroka@phys.ethz.ch}
%following the order of author 
%contribution{muSR}
%
%contribution{Sample}
%
%contribution{DFT}
%
%contribution{specific heat}
%
%contribution{characterization}
%
%contribution{specific heat}
%
%contribution{TDO}
%
%contribution{boss}
%
%%contribution{boss}
%
%contribution{boss}
%
%contribution{muSR + project coordination}

\begin{indented}
\item
\today
 %\now
\end{indented}

\begin{abstract}
In search of the origin of superconductivity in diluted rhenium
super\-conductors and their  
significantly enhanced 
$T_c$ compared 
to pure Be (0.026\,K), we investigated the intermetallic ReBe$_{22}$ 
compound, mostly by means of muon-spin rotation/relaxation ($\mu$SR). 
At a macroscopic level, its bulk superconductivity (with $T_c=9.4$\,K)   
was studied via electrical resistivity, magnetization, and 
heat-capacity measurements. The %low-temperature 
superfluid density, as determined from transverse-field 
$\mu$SR and electronic specific-heat 
measurements, suggest that ReBe$_{22}$ is a fully-gapped 
superconductor with some multigap features. 
The larger gap value, $\Delta_0^l=1.78$\,k$_\mathrm{B}T_c$, with a weight of 
almost 90\%, is slightly higher than that expected from the BCS theory 
in the weak-coupling case. 
The multigap feature, rather unusal for an almost 
elemental superconductor, is further supported by the field-dependent 
specific-heat coefficient, the temperature dependence of the upper 
critical field, as well as by electronic band-structure calculations. 
The absence of spontaneous magnetic fields below $T_c$, as 
determined from zero-field $\mu$SR measurements, indicates a 
preserved  time-reversal symmetry in the superconducting state of 
ReBe$_{22}$. In general, we find that a dramatic increase in the density of 
states at the Fermi level and an increase in the electron-phonon coupling 
strength, both contribute to the highly enhanced $T_c$ value of ReBe$_{22}$.
\end{abstract}

\vspace{2pc}
\noindent{\it Keywords}: Superconductivity, multigap, intermetallics, $\mu$SR, time-reversal symmetry

%\submitto{\NJP}
\maketitle

\section{\label{sec:Introduction}Introduction}

As one of the lightest elements, beryllium exhibits high-frequency 
lattice vibrations, a  
condition for achieving superconductivity (SC) with 
a sizeable critical temperature. Yet, paradoxically, its $T_c = 0.026$\,K 
is so low~\cite{Falge1967}, that its superconductivity is often overlooked. 
Clearly, $T_c$ is affected also by the electron-phonon coupling strength 
(typically large in elements with covalent-bonding tendencies) 
and the density of states (DOS) at the Fermi level $N(\epsilon_{\mathrm{F}})$ 
(rather low in pure Be). 
The latter depends on the details of crystal structure 
and on atomic volume, both effects being nicely illustrated by 
metal-hydride SCs under pressure (see, e.g., reference~\cite{Duan2017}).
In this regard, recently researchers could demonstrate a purely 
phonon-mediated 
superconductivity with $T_c$ up to 250\,K in actinium hydrides at 
200\,GPa~\cite{Semenok2018}.
The key insight of this work was the discovery of a link between 
chemical composition and superconductivity. Namely, that the 
superconductivity is more likely to occur in materials containing 
metal atoms that are close to populating a new electronic subshell, 
such as the $d_1$- (Sc, Y, La, and Ac)  
or $p_0$ (Be, Mg, and Ca) elements.
In these cases, the electronic structure becomes highly sensitive to 
the positions of the neighboring atoms~\cite{Slocombe2015}, resulting 
in stronger electron-phonon interactions and a higher $N(\epsilon_{\mathrm{F}})$. 
Based on this intuition, Be-rich alloys may achieve a $T_c$ much higher 
than elementary beryllium, a prediction which turns out to be 
true for ReBe$_{22}$~\cite{Bucher1967}, 
whose $T_c \sim 9.6$\,K is almost 400(!) times higher than that of Be. 
This is a remarkable result, deserving more attention and a detailed investigation of the ReBe$_{22}$ electronic properties.

ReBe$_{22}$ represents also a very interesting case in an entirely 
different aspect. Recently, a number of studies have shown that 
Re-based superconductors exhibit unconventional superconducting behaviour. 
For example, in non-centrosymmetric $\alpha$-Mn-type Re$T$ alloys 
($T$ = transition metal), the time-reversal symmetry (TRS) is broken, 
and the upper critical field is close to the Pauli 
limit~\cite{Wysokinski2019,Shang2018,TianReNb2018}. Surprisingly, our 
previous results show that, below $T_c$, even pure Re breaks TRS, thus behaving as an unconventional 
superconductor~\cite{TianReNb2018}. 
While binary Re-based superconductors have been investigated in both 
the full- (pure Re) and the intermediate Re limit (Re$T$), it is not 
clear if the unconventional behavior, in particular the TRS breaking, 
persists also in the dilute Re limit. With only 4\% of Re content, 
ReBe$_{22}$ is a good test case to verify such scenario.

In this paper, we report on an extensive study of the physical 
properties in the normal and superconducting state of ReBe$_{22}$, 
by means of electrical resistivity, magnetization, thermodynamic, 
and muon-spin relaxation ($\mu$SR) methods. In addition, we also 
present numerical density-functional-theory (DFT) band-structure 
calculations. ReBe$_{22}$ exhibits a fully-gapped, spin-singlet 
superconducting state with preserved TRS. Despite the very small amount 
of Re, the ReBe$_{22}$ alloy shows a remarkable increase in $T_c$ 
compared to its elementary constituents, which we mostly attribute to 
the significant increase of DOS at the Fermi level.

\section{Methods\label{sec:details}}
%
%==== figure =============================%
\begin{figure}[th]
	\centering
	\includegraphics[width=0.7\textwidth,angle=0]{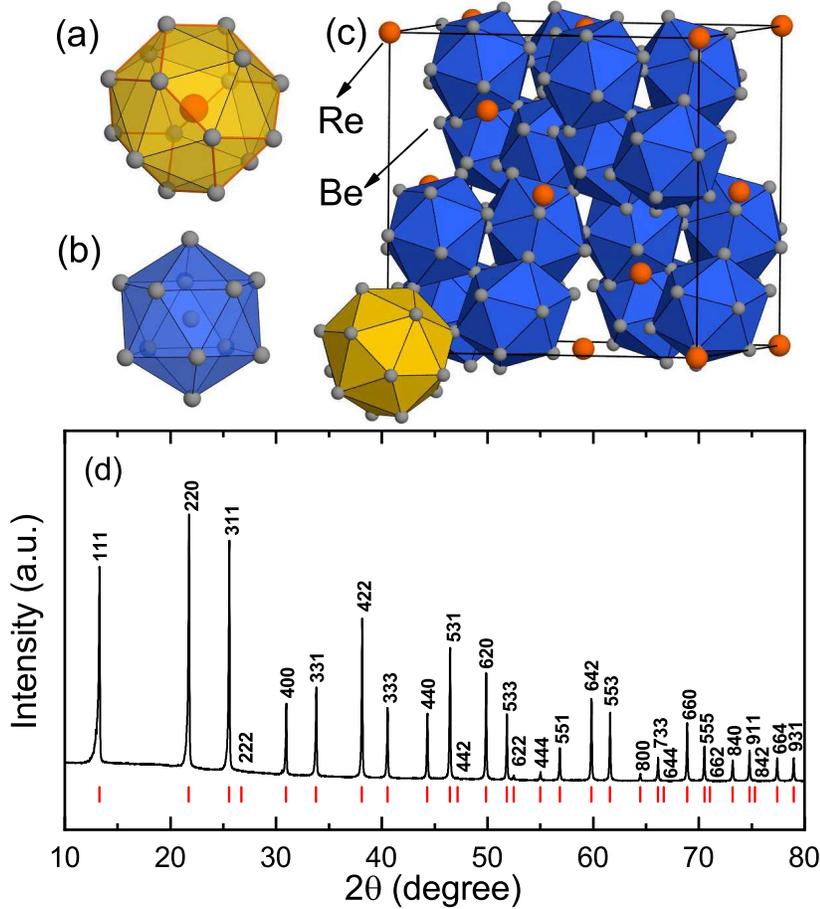}
	\vspace{-2ex}%
	\caption{\label{fig:structure} Crystal structure of ReBe$_{22}$. (a) ReBe$_{16}$ 
		units: the Be atoms (grey) around Re (red) occupy the vertices 
		of a truncated tetrahedron (i.e., a Friauf polyhedron), highlighted by red connection lines. Four additional Be atoms cap the hexagonal faces.
        (b) The distorted 
		Be$_{13}$ icosahedra consist exclusively of Be atoms. (c) Arrangement 
		of the ReBe$_{16}$ and Be$_{13}$ polyhedra in the unit cell. 
		For clarity, only one coordination polyhedron (yellow) around Re is drawn,
		the others being represented by red spheres. 
		(d) Powder x-ray diffraction pattern of ReBe$_{22}$. The vertical bars mark 
		the calculated Bragg-peak positions using the $Fd\bar{3}m$ space group.}
\end{figure}

Polycrystalline samples of ReBe$_{22}$ were prepared by arc melting of elementary 
Be (Heraeus, 99.9\%) and Re (Chem\-pur, 99.97\%) in an argon-filled glove box [MBraun, 
$p$(H$_{2}$O/O$_2$) $<$ 0.1\,ppm], dedicated to the handling of Be-containing 
samples \cite{LHS}. To compensate for the evaporation losses and to avoid the formation of spurious 
Re-Be binary phases, a small excess of beryllium was used.
Powder x-ray diffraction (XRD) measurements were performed on a Huber 
G670 image-plate Guinier camera (Ge-monochromator, Cu K$\alpha_1$ radiation). 
The lattice parameter of cubic ReBe$_{22}$ was determined from a least-squares 
fit to the experimental peak positions. The sample purity was
then checked by electron microscopy and energy-dispersive x-ray spectroscopy 
(EDX) on a JEOL JSM-6610 scanning electron microscope equipped with secondary 
electron-, electron backscatter-, 
and UltraDry EDS detectors (see figure~\ref{fig:edx} in the Appendix). 
Besides traces of elemental Be, no chemical impurities or secondary phases could 
be detected.

The magnetic susceptibility, electrical resistivity, and specific-heat measurements 
were performed on a 7-T Quantum Design Magnetic Property Measurement 
System (MPMS-7) and on a 14-T Physical Property Measurement System (PPMS-14) 
equipped with a $^3$He option. The $\mu$SR measurements were carried out 
at the GPS spectrometer of the Swiss muon source at Paul Scherrer Institut, 
Villigen, Switzerland~\cite{Amato2017}. The $\mu$SR data were analysed by 
means of the \texttt{musrfit} software package~\cite{Suter2012}.

The band structure of ReBe$_{22}$ was calculated by means of 
density-functional theory. Here we used the full-potential nonorthogonal local 
orbital code ($\texttt{FPLO}$)~\cite{Koepernik1999}.
To calculate the nonmagnetic band structure we employed the local-density 
approximation parametrized by the exchange-correlation potential of Perdew 
and Wang~\cite{Perdew1992}. The strong spin-orbit coupling of Re atoms was 
taken into account by performing full-relativistic calculations by solving the 
Dirac Hamiltonian with a generic potential. 

\section{\label{sec:results} Results and discussion}
\subsection{\label{ssec:structure} Crystal structure}
%
%=== end figure ==========================%
As shown in figure~\ref{fig:structure}, the complex intermetallic compound ReBe$_{22}$ 
adopts a cubic ZrZn$_{22}$-type structure with space group $Fd\overline{3}m$ 
(No.\ 227) and  $Z = 8$ formula units per cell. The lattice parameter 
$a$ = 11.5574(4)~\AA{}, determined from the XRD pattern [see 
figure~\ref{fig:structure}(d)], is consistent with the previously reported 
value~\cite{Sands1962}. No obvious impurity phases could be detected, 
indicating the high quality of the synthesized samples. The crystal 
structure can be visualized by means of two structural motifs. 
In the ReBe$_{16}$ motif [see figure~\ref{fig:structure}(a)], Re is 
coordinated by twelve Be atoms, lying 2.53~\AA{} apart at 
the vertices of a truncated tetrahedron, also known as a Friauf 
polyhedron. Four additional Be atoms lie atop the hexagonal faces of 
the truncated tetrahedron at a distance of 2.50\,\AA{} from the center.
A similar motif is also found in the NbBe$_{2}$ superconductor~\cite{Sands1959,Schirber1992}. 
As for the rest of Be atoms, these form distorted Be-centered Be$_{13}$ icosahedra, with the 
short interatomic distances 
ranging from 2.05 to 2.29~\AA{} [figure~\ref{fig:structure}(b)].  
Such Be-icosahedra represent the structural building blocks of the 
complex $M$Be$_{13}$ phases with a NaZn$_{13}$-type structure~\cite{Hidaka2017}. 

As shown in figure~\ref{fig:structure}(c), the ReBe$_{16}$ and Be$_{13}$ 
units are connected by sharing the polyhedra vertices. 
The arrangement of the two types of polyhedra within a unit cell can be described 
as hierarchically derived from the MgCu$_{2}$-type structure, where 
the Mg positions are occupied by ReBe$_{16}$ units and those of 
Cu by Be$_{13}$ icosahedra~\cite{Grin2013}. Both motifs, the truncated 
tetrahedron and the icosahedron, are found in the close-packed Laves phase 
structures. As a consequence of the high Be content, ReBe$_{22}$ features 
structural motifs  typically found in Be-rich intermetallic compounds,  
dominated by close-packing structures similar to that of 
hcp-Be~\cite{Samsonov1966,Jacobson2002}. Since the ratio of metallic 
radii [$r_\mathrm{Re}$/$r_\mathrm{Be}$ = 1.223] is close to the ideal 
value of 1.225, this facilitates the close-packing of unequal spheres in 
ReBe$_{22}$ and the accomodation of Re in the structure~\cite{Pauling1960,Stein2004}. 
The high packing fraction in this deltahedral structure is an important factor for 
the stabilization of this unusual stoichiometry,
also found in the isostructural MoBe$_{22}$ and WBe$_{22}$ compounds, 
both featuring similar ratios of radii~\cite{Paine1960}.

\subsection{\label{ssec:rho}Electrical resistivity}
%

%==== figure =============================%
\begin{figure}[th]
	\centering
	\includegraphics[width=0.8\textwidth,angle=0]{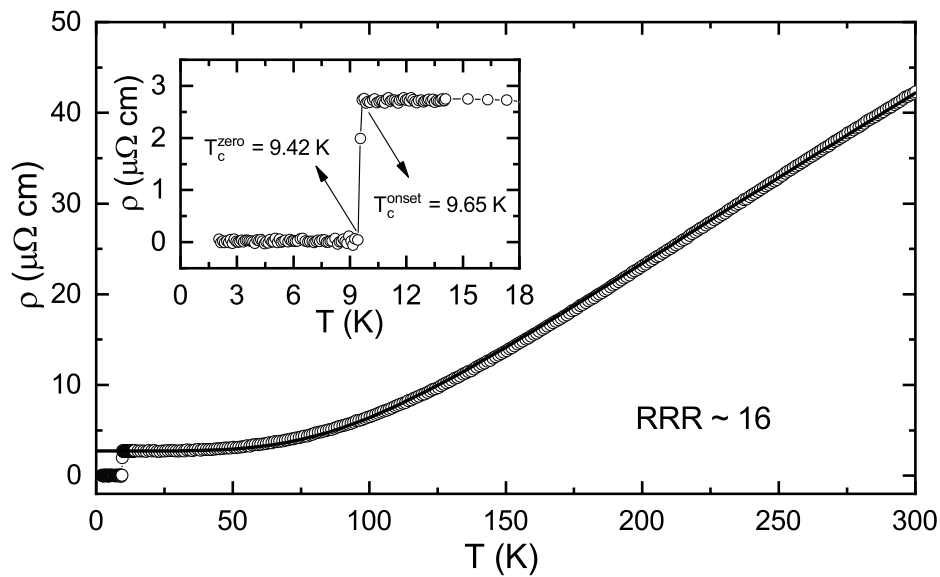}
	\vspace{-2ex}%
	\caption{\label{fig:Rho} Temperature dependence of the electrical 
		resistivity of ReBe$_{22}$. The solid line through the data is a fit to 
		equation (\ref{eq:rho_phonon}). The inset shows the data in the 
		low-temperature region, highlighting the superconducting transition. }
\end{figure}
%=== end figure ==========================%
%
%
The temperature dependence of the electrical resistivity $\rho(T)$ of ReBe$_{22}$
was measured in zero magnetic field from 300\,K down to 2\,K. As shown in 
figure~\ref{fig:Rho}, the resistivity exhibits metallic features down to base temperature, 
dropping to zero at the superconducting transition at $T_c^\mathrm{zero} = 9.42$\,K 
(see inset). 
Between $T_c$ and 300\,K the electrical resistivity  can be modelled by the 
Bloch-Gr\"{u}neisen (BG) formula~\cite{Bloch1930,Blatt1968}: 
%
%==== figure =============================%
\begin{figure}[th]
	\centering
	\includegraphics[width=0.7\textwidth,angle=0]{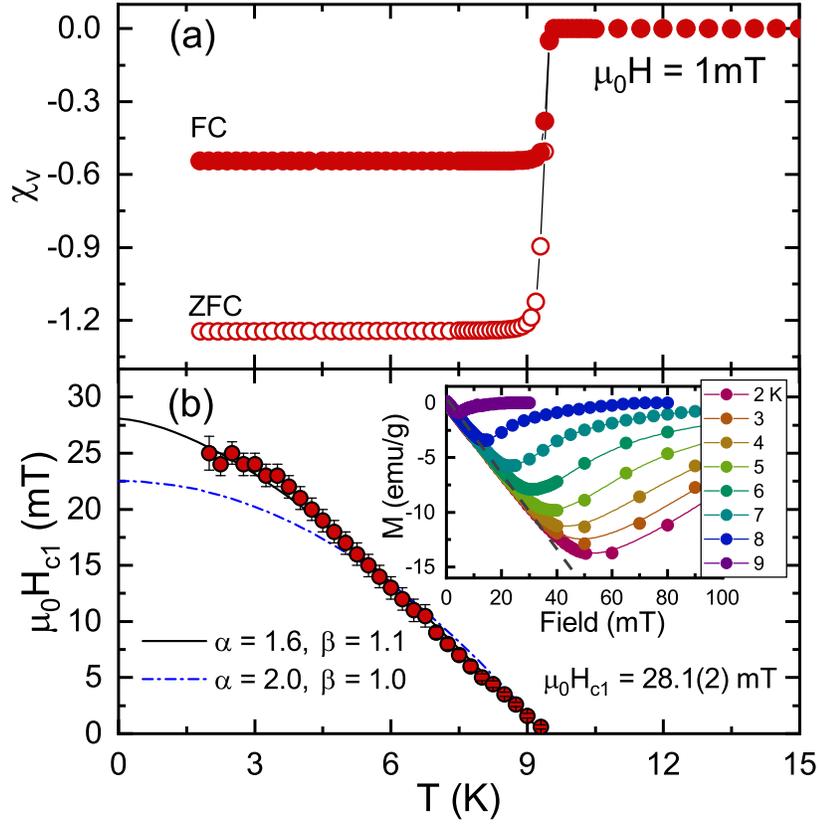}
	\vspace{-2ex}%
	\caption{\label{fig:Chi} (a) Temperature dependence of the magnetic susceptibility 
		of ReBe$_{22}$.  Data were collected in a 1-mT applied field using 
		both ZFC- and FC protocols. 
		(b) Estimated lower critical field $\mu_{0}H_{c1}$ as a function of 
		temperature. Lines are fits to the phenomenological model 
		$\mu_{0}H_{c1}(T) =\mu_{0}H_{c1}(0)[1-(T/T_{c})^\alpha]^\beta$. 
		The inset shows the field-dependent magnetization $M(H)$ recorded 
		at various temperatures up to $T_c$. For each temperature, 	the 
		lower critical field $\mu_{0}H_{c1}$ was determined as the value 
		where $M(H)$ starts deviating from linearity (dashed line).}
\end{figure}
%=== end figure ==========================%
%
\begin{equation}
\label{eq:rho_phonon}
\rho(T) = \rho_0 + 4A \left(\frac{T}{\Theta_\mathrm{D}^\mathrm{R}}\right)^5\int_0^{\frac{\Theta_\mathrm{D}^\mathrm{R}}{T}}\!\!\frac{z^2\mathrm{d}z}{(e^z-1)(1-e^{-z})}.
\end{equation}
Here, the first term $\rho_0$ is the residual resistivity due to the scattering 
of conduction electrons on the static defects of the crystal lattice, while the 
second term describes the electron-phonon scattering, with 
$\Theta_\mathrm{D}^\mathrm{R}$ being the characteristic (Debye) temperature 
and $A$ a coupling constant. The fit (black-line) in figure~\ref{fig:Rho} results in 
$\rho_0 = 2.72(5)$\,$\mu$$\Omega$cm, $A = 95(3)$\,$\mu$$\Omega$cm, 
and $\Theta_\mathrm{D}^\mathrm{R} = 590(5)$\,K. Such a large 
$\Theta_\mathrm{D}^\mathrm{R}$ value is consistent with the heat-capacity 
results (see below) and reflects the high frequency of phonons in ReBe$_{22}$. 
This is compatible with the high Debye temperature of elemental Be 
($\sim$ 1031\,K)~\cite{Chase1974}, in turn reflecting the small mass of 
beryllium atoms. 
A relatively large residual resistivity ratio [RRR $ = \rho(300\,\mathrm{K})/\rho_0 \sim 16$]   
and a sharp superconducting transition ($\Delta T = 0.23$\,K) 
both indicate a good sample quality. 

\subsection{\label{ssec:sus}Magnetization measurements}

The bulk superconductivity of ReBe$_{22}$ can be probed by magnetization measurements.
The temperature evolution of the magnetic susceptibility $\chi(T)$, 
measured at 1\,mT using both field-cooled (FC) and zero-field-cooled (ZFC) protocols, 
is shown in figure~\ref{fig:Chi}(a).
The splitting of the FC- and
ZFC-susceptibilities is typical of granular superconductors,
where the magnetic-field flux is trapped (in open holes) upon cooling the material in an applied field~\cite{Poole2014}.
The $\chi(T)$ curves show the onset of the 
superconducting transition at $T_c =9.50$\,K, in agreement with the values determined 
from electrical resistivity (figure~\ref{fig:Rho}) and heat capacity (see below). 
%%==== figure =============================%
\begin{figure}[th]
	\centering
	\includegraphics[width=0.8\textwidth,angle=0]{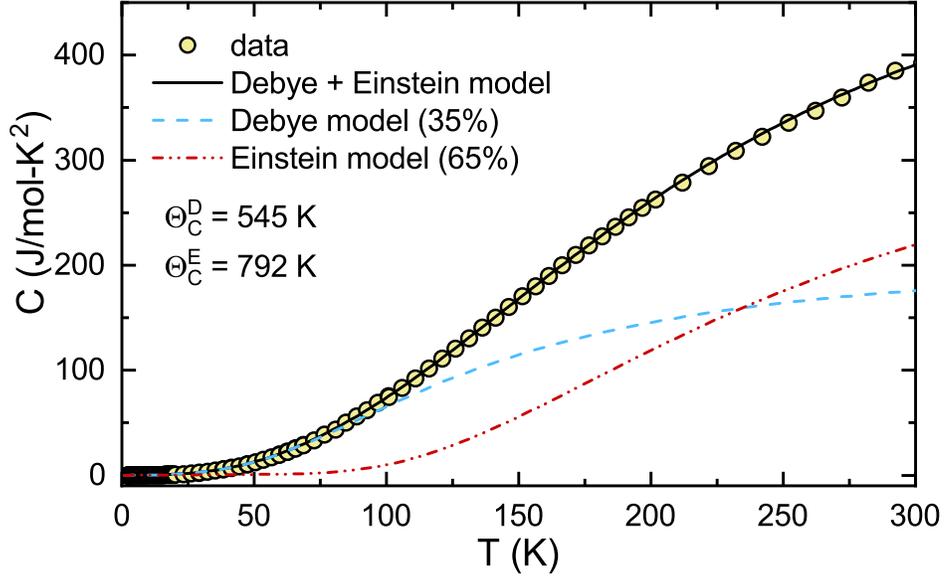}
	\vspace{-2ex}%
	\caption{\label{fig:Cp1}Temperature dependence of the heat capacity 
		measured in zero field between 2 and 300\,K. The solid line represents 
		a fit to a combined Debye and Einstein model, with the dashed- 
		and the dash-dotted lines referring to the two components.} 
\end{figure}
%=== end figure ==========================%
%

The field-dependent magnetization $M(H)$, measured at various 
temperatures (up to $T_c$), was used to determine the lower critical 
field $\mu_{0}H_{c1}$ of ReBe$_{22}$. As shown in the inset of figure~\ref{fig:Chi}(b), the $M(H)$ 
curves, recorded using a ZFC-protocol, exhibit the typical response expected 
for a type-II superconductor. The resulting $\mu_{0}H_{c1}$ vs.\ temperature 
data are summarized in figure~\ref{fig:Chi}(b) and the phenomenological 
model $\mu_{0}H_{c1}(T) =\mu_{0}H_{c1}(0)[1-(T/T_{c})^\alpha]^\beta$ 
was used to estimate $\mu_{0}H_{c1}(0)$. With $\alpha = 1.6$ and 
$\beta = 1.1$, the curve shown by a solid line in figure~\ref{fig:Chi}(b) 
gives a lower critical field $\mu_{0} H_{c1}(0) = 28.1(2)$\,mT. 
At the same time, the general model, with $\alpha = 2$ and  $\beta = 1$, 
shows a poor agreement with the experimental data. 
Since $\mu_{0}H_{c1}$ is proportional to the inverse-square of the 
magnetic penetration depth (see section~\ref{ssec:critical_field} and 
reference~\cite{Brandt2003}), 
a complex $T$-dependence of $\mu_{0}H_{c1}$ is 
indicative of multiband superconductivity in ReB$_{22}$.

\subsection{\label{ssec:Cp_zero} Specific heat}%%

The temperature dependence of the heat capacity $C(T)$ of ReBe$_{22}$ 
was also measured in zero-field conditions from 300\,K down to 2\,K. 
Although a single Debye- or Einstein model cannot  
describe the data, as shown in figure~\ref{fig:Cp1}, the normal-state $C(T)$ 
can be fitted by a combined Debye and Einstein model, with relative weights $x$ and $(1-x)$~\cite{Tari2003}:
\begin{equation}
\label{eq:Debye_Einstein}
C(T) = \gamma_\mathrm{n} T + n xC_\mathrm{D}(T) + n (1-x)C_\mathrm{E}(T).
\end{equation}
The number of atoms per ReBe$_{22}$ formula-unit ($n = 23$) is 
considered in the above equation. The first term represents 
the electronic specific heat, which can be determined from the low-$T$ 
heat-capacity data (see below). The second and the third terms 
represent the acoustic- and optical phonon-mode contributions, 
described by the Debye and Einstein model, respectively~\cite{Tari2003}:

\begin{equation}
\label{eq:Debye}
C_\mathrm{D}(T) =  9R \left(\frac{T}{\Theta_\mathrm{D}^\mathrm{C}}\right)^3\int_0^{\frac{\Theta_\mathrm{D}^\mathrm{C}}{T}}\!\!\frac{z^{4}e^z\mathrm{d}z}{(e^z-1)^2},
\end{equation}   
\begin{equation}
\label{eq:Einstein}
C_\mathrm{E}(T) = 3R \left(\frac{\Theta_\mathrm{E}^\mathrm{C}}{T}\right)^2\frac{\mathrm{exp}(\Theta_\mathrm{E}^\mathrm{C}/T)}{[\mathrm{exp}(\Theta_\mathrm{E}^\mathrm{C}/T)-1]^2}.
\end{equation}
Here $\Theta_\mathrm{D}^\mathrm{C}$ and $\Theta_\mathrm{E}^\mathrm{C}$ 
are the Debye and Einstein temperatures, while $R = 8.314$\,J/mol-K is the molar 
gas constant. 
The best fit curve (solid line in figure~\ref{fig:Cp1}) is obtained for 
$\Theta_\mathrm{D}^\mathrm{C} = 545(5)$\,K and $\Theta_\mathrm{E}^\mathrm{C} = 792(5)$\,K,  
with $x = 0.35(2)$. The resulting Debye temperature is comparable to that derived 
from electrical resistivity data (see figure~\ref{fig:Rho}). 

%==== figure =============================%
\begin{figure}[th]
	\centering
	\includegraphics[width=0.8\textwidth,angle=0]{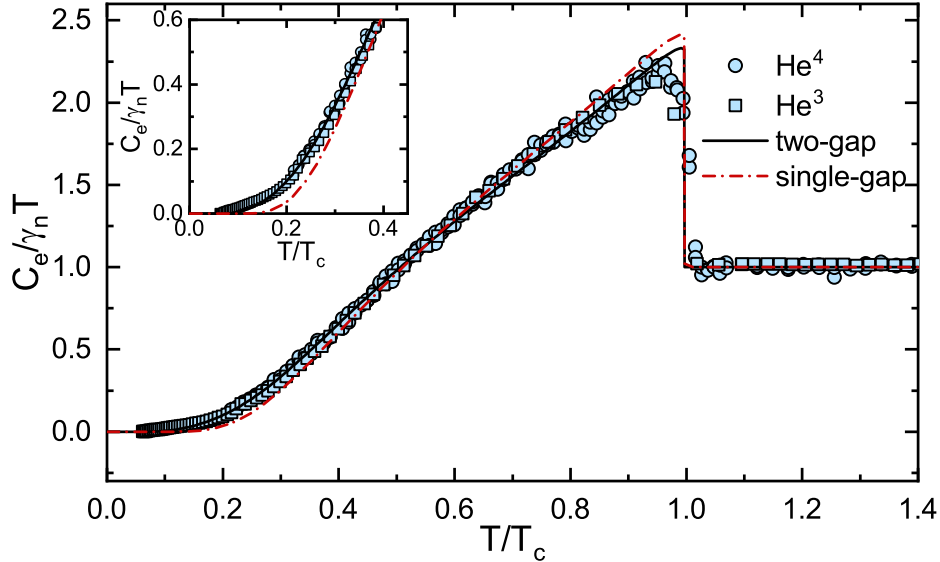}
	\vspace{-2ex}%
	\caption{\label{fig:Cp2}Normalized electronic specific heat 
		$C_\mathrm{e}/\gamma_n T$ of ReBe$_{22}$ as a function of $T$/$T_c$, 
		measured using $^4$He- (circles) and $^3$He cooling (squares). 
	%	Upper inset: specific heat $C/T$ vs.\ $T^2$; the dashed-line is a fit 
		%to $C/T = \gamma_n + \beta T^2 + \delta T^4$. 
		Inset: low-$T$ region of $C_\mathrm{e}/\gamma_n T$. The solid- 
		and the dash-dotted lines %in the main panel 
		represent the electronic 
		specific heat calculated by considering a fully-gapped $s$-wave model 
		with two- and one gap, respectively.} 
\end{figure}
%=== end figure ==========================%

The low-$T$ specific-heat data were further analyzed, since they can  
offer valuable insight into the superconducting properties of ReBe$_{22}$ through 
the evaluation of the quasiparticle DOS at the Fermi level. 
As shown in  figure~\ref{fig:Cp2}, the sharp specific-heat jump at $T_c$ 
again indicates a bulk superconducting transition and a good sample quality. 
The electronic specific heat $C_\mathrm{e}$/$T$ was obtained by 
subtracting the phonon contribution from the experimental data.  
The DOS at the Fermi level $N(\epsilon_\mathrm{F})$ 
can be  evaluated  from the expression 
$N(\epsilon_\mathrm{F}) = 3\gamma_\mathrm{n}/(2\pi^2 k_\mathrm{B}^2) = 3.25(3)$\,states/eV-f.u.\ 
(accounting for spin de\-ge\-ne\-ra\-cy)~\cite{Kittel2005}, where 
$k_\mathrm{B}$ is the Boltzmann constant and $\gamma_\mathrm{n} = 15.3(2)$\,mJ/mol-K$^2$ is the electronic specific-heat coefficient. The electron-phonon coupling 
constant $\lambda_\mathrm{ep}$, a measure of the attractive interaction 
between electrons due to phonons, was estimated from the 
$\Theta_\mathrm{D}^\mathrm{C}$ and $T_c$ values by means of the semi-empirical 
McMillan formula~\cite{McMillan1968}:
\begin{equation}
\label{eq:lambda_ep}
\lambda_\mathrm{ep}=\frac{1.04+\mu^{\star}\,\mathrm{ln}(\Theta_\mathrm{D}/1.45\,T_c)}{(1-0.62\,\mu^{\star})\mathrm{ln}(\Theta_\mathrm{D}/1.45\,T_c)-1.04}.
\end{equation}
The Coulomb pseudo-potential $\mu^{\star}$ was fixed to 0.13, a typical 
value for metallic samples. From equation (\ref{eq:lambda_ep}) we obtain 
$\lambda_\mathrm{ep} = 0.64(1)$ for ReBe$_{22}$, almost \emph{three times 
	larger} than the  reported value for elemental Be (0.21)~\cite{Sklyadneva2005}. 
By using this value, finally, the band-structure density of states 
$N_\mathrm{band}(\epsilon_\mathrm{F})$ can be estimated from the relation 
$N_\mathrm{band}(\epsilon_\mathrm{F}) = N(\epsilon_\mathrm{F})/(1 + 
\lambda_\mathrm{ep}$)~\cite{Kittel2005}, which gives 
$N_\mathrm{band}(\epsilon_\mathrm{F})$ = $1.98(2)$\,states/eV-f.u.

After subtracting the phonon contribution from the %measured 
specific-heat data, the electronic specific heat divided by the electronic 
specific-heat coefficient, i.e., $C_\mathrm{e} / \gamma_\mathrm{n} T$, 
is obtained %as shown in the 
(main panel in figure~\ref{fig:Cp2}). 
The temperature-dependent superconducting-phase contribution to the 
entropy was calculated by means of the BCS expression~\cite{Tinkham1996}:
%
%==== figure =============================%
\begin{figure}[th]
	\centering
	\includegraphics[width=0.7\textwidth,angle=0]{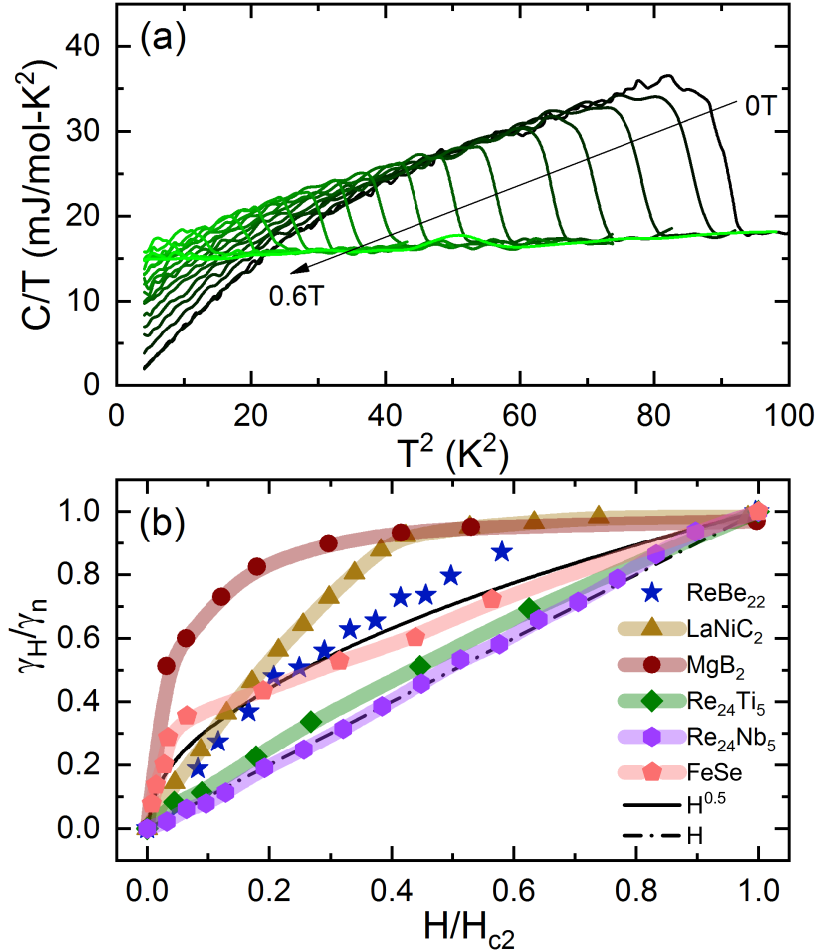}
	\vspace{-2ex}%
	\caption{\label{fig:Cp3}(a) Specific heat of ReBe$_{22}$ as a function of 
		$T^2$, measured under increasing magnetic fields (up to 0.6\,T). 
		(b) Normalized specific-heat coefficient $\gamma_\mathrm{H}$/$\gamma_\mathrm{n}$ 
		vs.\ the reduced magnetic field value $H/H_\mathrm{c2}(0)$. $\gamma_\mathrm{H}$ 
		is estimated by extrapolating the data in (a) to zero temperature. The 
		dash-dotted line indicates a linear dependence as predicted for an $s$-wave 
		gap structure, the solid line represents the dependence expected for an anisotropic gap 
		or a gap with nodes, e.g., $d$-wave. The data for the reported samples 
		were adopted from the references~\cite{Shang2018,TianReNb2018,Chen2013,Chen2017,Bouquet2001a}.} 
\end{figure}
%=== end figure ==========================%
%
\begin{equation}
\label{eq:entropy}
S(T) = -\frac{6\gamma_\mathrm{n}}{\pi^2 k_\mathrm{B}} \int^{\infty}_0 [f\mathrm{ln}f+(1-f)\mathrm{ln}(1-f)]\,\mathrm{d}\epsilon,
\end{equation}
where $f = (1+e^{E/k_\mathrm{B}T})^{-1}$ is the Fermi function and 
$E(\epsilon) = \sqrt{\epsilon^2 + \Delta^2(T)}$ is the excitation energy 
of quasiparticles, with $\epsilon$ the electron energies measured relative 
to the chemical potential (Fermi energy)~\cite{Tinkham1996,Padamsee1973}. 
Here $\Delta(T) = \Delta_0 \mathrm{tanh} \{ 1.82[1.018(T_\mathrm{c}/T-1)]^{0.51} \}$ 
\cite{Carrington2003}, with $\Delta_0$ the gap value at zero temperature. 
The temperature-dependent electronic specific heat in the superconducting 
state can be calculated from $C_\mathrm{e} =T \frac{dS}{dT}$. 
The dash-dotted line in figure~\ref{fig:Cp2} represents a fit with an 
$s$-wave model with a single gap $\Delta_0 = 1.40(1)$\,meV. While this
reproduces well the experimental data in the %temperature 
$0.4 < T/T_c < 0.8$ range, out of it the single-gap model clearly 
deviates from the data (see lower inset).
On the contrary, the two-gap model exhibits a better agreement, both at 
low temperatures as well as near $T_c$. The solid line in figure~\ref{fig:Cp2} 
is a fit to the two-gap $s$-wave model, known also as $\alpha$ model~\cite{Bouquet2001}: 
\begin{equation}
\label{eq:two_gap}
C_e(T)/T = wC_e^{\Delta^s}(T)/T + (1-w)C_e^{\Delta^l}(T)/T.
\end{equation}
Here each term represents 
a single-gap specific-heat contribution, with $\Delta^s$ the small- and $\Delta^l$ 
the large gap, and $w$ the relative weight. 
The two-gap model gives $w = 0.13$, $\Delta_0^s = 0.68(1)$\,meV 
and $\Delta_0^l = 1.43(1)$\,meV, 
with both superconducting gap values being consistent 
with the $\mu$SR results (see figure~\ref{fig:lambda}). 
In addition, the larger gap is comparable to the weak-coupling BCS 
value (1.4\,meV), indicating weakly-coupled superconducting pairs in 
ReBe$_{22}$.
The specific-heat discontinuity at $T_c$, i.e., $\Delta C/\gamma_\mathrm{n} T_{c} = 1.24$, 
is smaller than the BCS value of 1.43. There are two possibilities for such 
a reduced specific-heat discontinuity, 
despite a good sample quality and full superconducting volume fraction: 
i) gap anisotropy, including a nodal gap, as observed in some heavy-fermion 
superconductors or in Sr$_2$RuO$_4$~\cite{Movshovich2001,Mackenzie2003}, 
or ii) multiband superconductivity, as e.g., in MgB$_2$ or LaNiGa$_2$~\cite{Bouquet2001a,Weng2016}. 
Due to a highly-symmetric crystal structure and to a lack of gap nodes (see below), 
only the second scenario is applicable to the ReBe$_{22}$ case.

The multiband superconductivity of ReBe$_{22}$ can be inferred also 
from the field dependence of the electronic specific heat coefficient 
$\gamma_\mathrm{H}$. As shown in figure~\ref{fig:Cp3}(a), 
at a given applied field, $\gamma_\mathrm{H}$ is obtained as the 
linear extrapolation of $C/T$ vs.\ $T^2$ (in the superconducting phase) 
to zero temperature. The dependence of the normalized $\gamma_\mathrm{H}/\gamma_\mathrm{n}$ %\tcr{value} 
vs.\ the reduced magnetic field $H/H_\mathrm{c2}(0)$ is shown 
in figure~\ref{fig:Cp3}(b) (here $\gamma_\mathrm{n}$ is the zero-field 
normal-phase value). Note that, the field dependence of $\gamma_\mathrm{H}/\gamma_\mathrm{n}$ at 2\,K 
exhibits similar features to that evaluated at zero temperature.
Due to the multiband effects, it is difficult to describe the field dependence 
of $\gamma_\mathrm{H}$ in ReBe$_{22}$ with a simple formula. As can be 
seen in figure~\ref{fig:Cp3}(b), $\gamma_\mathrm{H}(H)$ clearly deviates 
from the linear field dependence (dash-dotted line) expected for 
single-gap BCS superconductors~\cite{Caroli1964}, 
or from the square-root dependence $\sqrt{H}$ (solid line) expected for 
nodal superconductors~\cite{Volovik1993,Wen2004}. In fact, ReBe$_{22}$ exhibits 
similar features to other multiband superconductors, as e.g., LaNiC$_2$~\cite{Chen2013}, 
FeSe~\cite{Chen2017}, and MgB$_2$~\cite{Bouquet2001a} (the latter being 
a prototypical two-gap superconductor), although the slopes 
of $\gamma_\mathrm{H}(H)$ close to $H=0$ are different.

\begin{figure}[htp]
	\centering
	\includegraphics[width=0.8\textwidth,angle= 0]{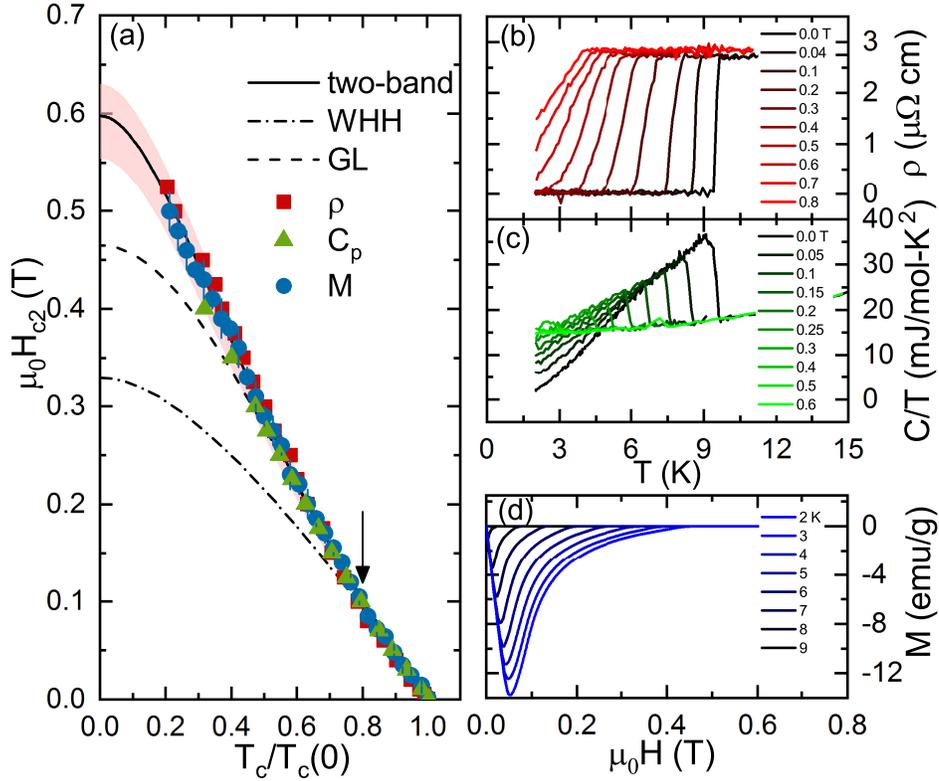}
	\caption{\label{fig:Hc2_determ}(a) Upper critical field $\mu_{0}H_{c2}$ vs.\ 
		reduced transition temperature $T_c/T_c(0)$ for ReBe$_{22}$. The $T_c$ values 
		were determined from temperature-dependent electrical resistivity $\rho(T,H)$ (b) 
		and specific-heat $C(T,H)/T$ data (c) at various applied fields, and from 
		field-dependent magnetization $M(H,T)$ (d) at different temperatures. For the 
		$\rho(T,H)$ measurements, $T_c$ was defined as the onset of zero resistivity. 
		Three different models, including two-band (solid line), WHH (dash-dotted line), and 
		GL model (dashed line), are shown in (a). In case of the WHH model, the spin-orbit 
		scattering was neglected. The shadowed region indicates the upper and lower 
		$H_{c2}$ limits, as determined using the two-band model.}
\end{figure}
%=== end figure ==========================%
%

\subsection{\label{ssec:critical_field}  Upper critical field}  

The upper critical field $\mu_0$$H_{c2}$ of ReBe$_{22}$ was determined 
via temperature-dependent resistivity $\rho(T, H)$ and specific heat $C(T, H)/T$ 
measurements at various applied magnetic fields, as well as from the 
field-dependent magnetization $M(H, T)$ at various temperatures. 
The derived $\mu_0H_{c2}$ values as a function of the reduced temperature 
$T_c$/$T_c$(0) are summarized in figure~\ref{fig:Hc2_determ}(a). Upon applying a magnetic field, the superconducting transition in both $\rho(T)$ 
and specific heat $C(T)$/$T$ data shifts towards lower temperatures [see figures~\ref{fig:Hc2_determ}(b)-(c)]. 
In the $M(H, T)$ case, the diamagnetic signal disappears once the applied 
magnetic field exceeds the upper critical field [figure~\ref{fig:Hc2_determ}(d)]. 
The $\mu_0 H_{c2}$ values determined using different techniques are highly consistent. 
The temperature dependence of $\mu_0 H_{c2}(T)$ was analyzed by means 
of three different models, i.e., a Ginzburg-Landau (GL)~\cite{Zhu2008}, 
a Werthamer-Helfand-Hohenberg (WHH)~\cite{Werthamer1966}, and a 
two-band (TB)~\cite{Gurevich2003} model. As can be seen in 
figure~\ref{fig:Hc2_determ}(a), at low fields, both GL and WHH models 
reproduce very well the experimental data. However, at higher magnetic 
fields, both models deviate significantly from the experimental data, 
providing underestimated $\mu_0 H_{c2}(0)$ values.

%model $\mu_0 H_{c2}(T) = \mu_0 H_{c2}(0)(1-t^2)/(1+t^2)$, where 
%$t = T/T_{c}$ is the normalized temperature (see, e.g., \cite{Zhu2008}). 
%Such model follows from the Ginzburg-Landau (GL) relation between 
%$H_{c2}$ and the coherence length $\xi$ (see below), by assuming 
%$\xi(t) \propto \sqrt{(1+t^2)/(1-t^2)}$. Although the GL theory is 
%strictly valid near $T_c$, the above relation has been proved to be satisfied 
%in a wider temperature range.
%the Ginzburg-Landau (GL) 
%model $\mu_0 H_{c2}(T) = \mu_0 H_{c2}(0)(1-t^2)/(1+t^2)$, where 
%$t = T/T_{c}$ is the normalized temperature. 
%The solid lines in figure~\ref{fig:Hc2_determ}(d) are fits to the GL model, 
%which gives $\mu_0 H_{c2}^\mathrm{GL}(0) = 0.94(1)$\,T and 1.09\,T for 
%specific heat and magnetization, and for electrical resistivity data, 
%respectively. For a comparison, we estimated the upper critical field 
%also by means of the Werthamer-Helfand-Hohenberg (WHH) model~\cite{Werthamer1966}. 
%The dashed-lines in figure~\ref{fig:Hc2_determ}(d) are fits to an WHH model 
%without spin-orbital scattering and give $\mu_0 H_{c2}^\mathrm{WHH}(0) = 0.76(1)$\,T 
%and 0.83(1)\,T for specific heat and magnetization, and for electrical 
%resistivity data, respectively. 

A positive curvature of $\mu_0 H_{c2}(T)$ near $T_c$ is considered a 
typical feature of multiband superconductors, as e.g., MgB$_2$~\cite{Muller2001,Gurevich2004}.  
It reflects the gradual suppression of the small superconducting gap 
with increasing magnetic field, as evidenced also by the specific-heat data shown in figure~\ref{fig:Cp3}.
The arrow in figure~\ref{fig:Hc2_determ}(a)
identifies the small kink in $\mu_0 H_{c2}(T)$ close to 0.1\,T, here considered to coincide with the field 
value which suppresses the small superconducting gap. Also $\gamma_\mathrm{H}$ 
changes its slope near this critical field [corresponding to $H/H_{c2} = 0.17$ 
in figure~\ref{fig:Cp3}(b)]. The remarkable agreement of the two-band model 
with the experimental data across the full temperature range is clearly seen in 
figure~\ref{fig:Hc2_determ}(a), from which we find $\mu_0 H_{c2}^\mathrm{TB}(0) = 0.60(3)$\,T.

The superconducting coherence length $\xi$ can be calculated from $\xi$ =  $\sqrt{\Phi_0/2\pi\,H_{c2}}$, 
where $\Phi_0 = 2.07 \times 10^{-3}$\,T~$\mu$m$^{2}$ is the quantum of magnetic flux. 
With a bulk $\mu_{0}H_{c2}(0) = 0.60(3)$\,T, the calculated $\xi(0)$ is 23(1)\,nm. 
The lower critical field $\mu_{0}H_{c1}$ is related to the magnetic penetration 
depth $\lambda$ and the coherence length $\xi$ via $\mu_{0}H_{c1} = (\Phi_0 /4 \pi \lambda^2)[$ln$(\kappa)+ 0.5]$, 
where $\kappa$ = $\lambda$/$\xi$ is the GL parameter~\cite{Brandt2003}.
By using $\mu_{0}H_{c1} = 28.1$\,mT and $\mu_{0}H_{c2} = 0.60(3)$\,T, 
the resulting magnetic penetration depth $\lambda_\mathrm{GL}$ = 109(1)\,nm, 
is comparable to 87(1)\,nm (40\,mT) and 104(1)\,nm (120\,mT), the 
experimental values evaluated from TF-$\mu$SR data (see section~\ref{ssec:TF_muSR}). 
A GL parameter $\kappa \sim 4.7(3)$, much
larger than the threshold value of  $1/\sqrt{2}$, 
clearly indicates that ReBe$_{22}$ is a type-II superconductor.

\subsection{\label{ssec:TF_muSR} Transverse-field $\mu$SR}
%=== begin figure ==========================%

%=== begin figure ==========================%
\begin{figure}[!thp]
	\centering
	\includegraphics[width=0.80\textwidth,angle= 0]{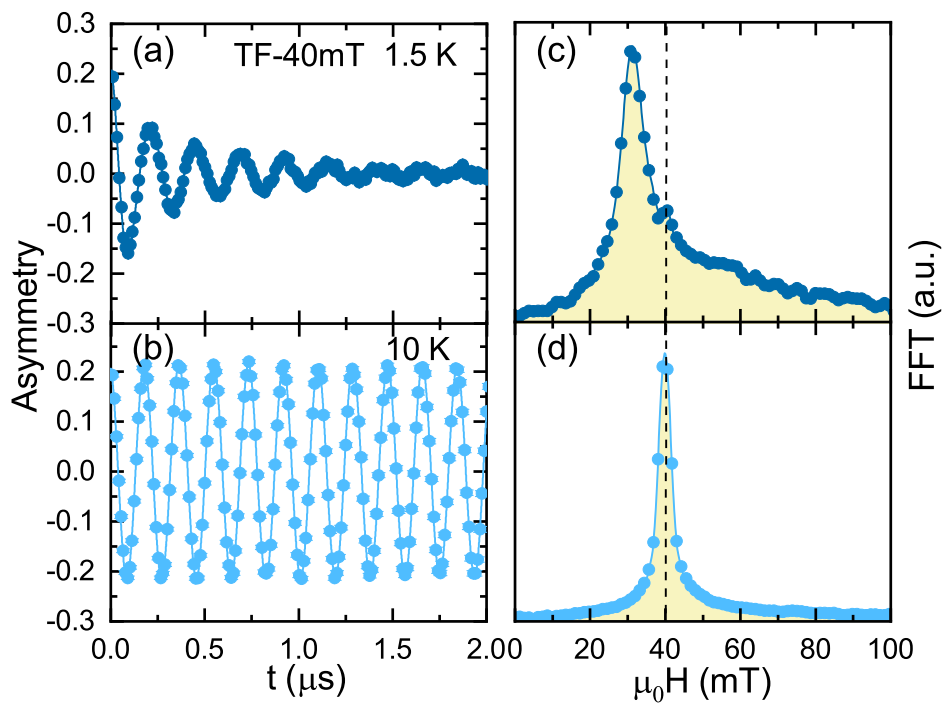}
	\caption{\label{fig:TF-muSR_T}TF-$\mu$SR time spectra collected
		at 1.5\,K (a) and 10 \,K (b) in an applied field of 40\,mT, with the 
		respective Fourier transforms being shown in (c) and (d). Solid lines are fits to equation (\ref{eq:TF_muSR})  
		using three oscillations. 
		The dashed vertical line indicates the applied magnetic field. 
		Note the clear diamagnetic shift below $T_{c}$ in panel (c). }
\end{figure}
%=== end figure ==========================% 

\begin{figure}[!thp]
	\centering
	\includegraphics[width=0.8\textwidth,angle= 0]{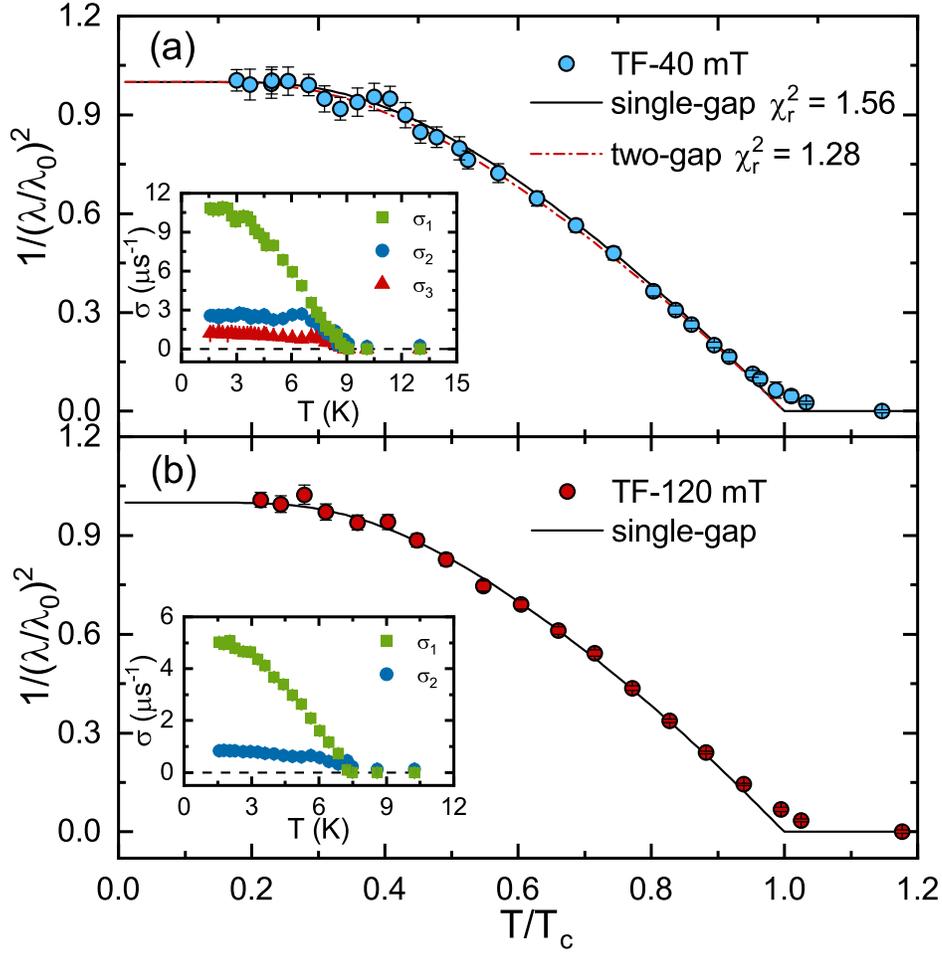}
	\caption{\label{fig:lambda} Superfluid density vs.\ temperature, as 
		determined from TF-$\mu$SR measurements in an applied magnetic field 
		of 40\,mT (a) and 120\,mT (b). The insets show the temperature dependence 
		of the muon-spin relaxation rate $\sigma(T)$. While three components 
		are required to describe the TF-40\,mT data, only two components 
		are necessary in the TF-120\,mT case.
		Lines represent fits to a fully-gapped $s$-wave 
		model with either two- (dash-dotted) or a single SC gap (solid).}
\end{figure}

$\mu$SR measurements in an applied transverse field (TF) were 
carried out to investigate the superconducting properties of ReBe$_{22}$ 
at a microscopic level. Preliminary field-dependent $\mu$SR depolarization-rate 
measurements at 1.5\,K were carried out to determine the optimal field value 
for the temperature-dependent study (see figure~\ref{fig:TF-muSR_H} in Appendix). 
To track the additional field-distribution broadening due to the flux-line-lattice 
(FLL) in the mixed superconducting state, the magnetic field was applied in 
the normal state, prior to cooling the sample below $T_c$. 
After the field-cooling protocol, which ensures an almost ideal FLL 
even in case of pinning effects, the TF-$\mu$SR measurements 
were performed at various temperatures upon warming.
Figures~\ref{fig:TF-muSR_T}(a) and (b) show two representative TF-$\mu$SR 
time-spectra collected in the superconducting (1.5\,K) and the normal state (10\,K) 
in an applied field of 40\,mT at the GPS spectrometer. 
The enhanced depolarization rate below $T_c$ reflects the inhomogeneous 
field distribution due to the FLL, causing an additional distribution 
broadening in the mixed state [see figure~\ref{fig:TF-muSR_T}(c)].
The $\mu$SR spectra can be modelled by the following expression: 
\begin{equation}
\label{eq:TF_muSR}
A_\mathrm{TF}(t) = \sum\limits_{i=1}^n A_i \cos(\gamma_{\mu} B_i t + \phi) e^{- \sigma_i^2 t^2/2} +
A_\mathrm{bg} \cos(\gamma_{\mu} B_\mathrm{bg} t + \phi).
\end{equation}
Here $A_i$ and $A_\mathrm{bg}$ represent the initial muon-spin 
asymmetries for muons implanted in the sample and sample holder, respectively, 
with the latter not undergoing any depolarization. $B_i$ and $B_\mathrm{bg}$ are the local fields sensed by implanted muons in the sample and sample holder, $\gamma_{\mu} = 2\pi \times 135.53$\,MHz/T is the muon gyromagnetic ratio, $\phi$ is a shared initial phase, and $\sigma_i$ is a Gaussian relaxation rate of the $i$th component. The number of required components is material dependent, generally in the 
$1 \leq n \leq 5$ range.  For superconductors with a large $\kappa$ ($\gg 1$), 
the magnetic penetration depth is much larger than the coherence length. 
Hence, the field profiles of each fluxon overlap strongly, implying 
a narrow field distribution. Consequently, a single-oscillating component 
is sufficient to describe $A(t)$, as e.g., in Re$T$~\cite{Shang2018,TianReNb2018} or 
Mo$_3$Rh$_2$N~\cite{Shang2018Mo}. In case of a small $\kappa$ 
($\gtrsim 1/\sqrt{2}$), as e.g., in ReBe$_{22}$, 
the magnetic penetration depth is comparable to the coherence length. 
The rather small $\lambda$ implies fast-decaying fluxon field profiles and 
a broad field distribution, in turn requiring multiple oscillations to 
describe $A(t)$~\cite{Maisuradze2009}.
The fast-Fourier-transform (FFT) spectra of the
TF-$\mu$SR datasets at 1.5\,K and 10\,K 
are shown in figures~\ref{fig:TF-muSR_T}(c) and (d). The solid lines 
represent fits to equation (\ref{eq:TF_muSR}) using three oscillations 
(i.e., $n = 3$) in the superconducting state and one oscillation in the 
normal state. The TF-$\mu$SR spectra collected at 120\,mT require only two 
oscillations (i.e., $n = 2$), indicating a narrower field distribution compared to the 
40-mT case. The derived Gaussian relaxation rates as a function 
of temperature are summarized in the insets of figure~\ref{fig:lambda}. 
%
%=== begin figure ==========================%
%=== end figure ==========================%  

%
Above $T_c$, the relaxation rate is small and temperature-independent, but 
below $T_c$ it starts to increase due to the onset of the FLL and the 
increase in superfluid density.
In case of multi-component oscillations, 
the first-term in equation (\ref{eq:TF_muSR}) describes the field distribution 
as the sum of $n$ Gaussian relaxations~\cite{Maisuradze2009}:
\begin{equation}
\label{eq:TF_muSR_2}
P(B) = \gamma_{\mu} \sum\limits_{i=1}^n \frac{A_i}{\sigma_i} \mathrm{exp}\left[-\frac{\gamma_{\mu}^2(B-B_i)^2}{2\sigma_i^2}\right].
\end{equation}
Then, the first- and the second moment of the field distribution 

can be calculated by: 
\begin{equation}
%\begin{aligned}
\label{eq:1st_moment}
	\langle B \rangle =  \sum\limits_{i=1}^n \frac{A_i B_i}{A_\mathrm{tot}},\quad \mathrm{and} \\
\end{equation}
\begin{equation}
\label{eq:2nd_moment}
 \langle B^2 \rangle = \frac{\sigma_\mathrm{eff}^2}{\gamma_\mu^2} = \sum\limits_{i=1}^n \frac{A_i}{A_\mathrm{tot}}\left[\frac{\sigma_i^2}{\gamma_{\mu}^2} - \left(B_i - \langle B \rangle\right)^2\right],
%\end{aligned}
\end{equation}
where $A_\mathrm{tot} = \sum_{i=1}^n A_i$. 
The superconducting Gaussian 
relaxation rate related to the FLL ($\sigma_\mathrm{FLL}$) can be 
extracted by subtracting the nuclear contribution according to 
$\sigma_\mathrm{FLL} = \sqrt{\sigma_\mathrm{eff}^{2} - \sigma^{2}_\mathrm{n}}$, 
where $\sigma_\mathrm{n}$ is the nuclear relaxation rate. 
The superconducting gap value and its symmetry can be investigated 
by measuring the temperature-dependent $\sigma_\mathrm{FLL}$$(T)$, 
which is directly related to the magnetic penetration depth and thus the 
superfluid density ($\sigma_\mathrm{FLL} \propto 1/\lambda^2$).

Since the upper critical field of ReBe$_{22}$ is relatively 
small (600\,mT) compared to the applied fields used in the TF-$\mu$SR 
study (40 and 120\,mT), the effects of the overlapping vortex cores 
with increasing field ought to be considered when extracting the magnetic 
penetration depth $\lambda$ from $\sigma_\mathrm{FLL}$. 
For ReBe$_{22}$, $\lambda$ was calculated by means 
of~\cite{Brandt2003,Barford1988}: 
\begin{equation}
\label{eq:TF_muSR_H}
\sigma_\mathrm{FLL} = 0.172 \frac{\gamma_{\mu} \Phi_0}{2\pi}(1-h)[1+1.21(1-\sqrt{h})^3]\lambda^{-2}, 
\end{equation} 
where $h = H_\mathrm{appl}/H_\mathrm{c2}$, with $H_\mathrm{appl}$ the 
applied magnetic field. The above expression is valid for type-II 
superconductors with $\kappa \ge 5$ in the $0.25/\kappa^{1.3} \lesssim h \le$ 1 
field range. With $\kappa \sim 4.7$ and $h = 0.067$ (TF-50\,mT) and 0.2 
(TF-120\,mT), ReBe$_{22}$ fulfills the above condition. Note that, in 
the above expression, only the absolute value of the penetration depth, 
but not its temperature dependence is related to the $h$ value. 
By using equation (\ref{eq:TF_muSR_H}), we calculated the inverse-square of 
the magnetic penetration depth, which is proportional to the superfluid 
density, i.e., $\lambda^{-2}(T) \propto \rho_\mathrm{sc}(T)$. As can be seen in 
figure~\ref{fig:lambda}, below $T_c/3$, $\rho_\mathrm{sc}(T)$ is 
practically independent of temperature, in agreement with the 
specific-heat results shown in figure~\ref{fig:Cp3}, once more 
indicating a nodeless superconductivity in ReBe$_{22}$. 
$\rho_\mathrm{sc}(T)$ was further analysed by means of a two-gap 
$s$-wave model, previously applied to the well-established two-gap 
superconductor MgB$_2$~\cite{Carrington2003,Nieder2002}. In general, 
the superfluid density can be described by:
\begin{equation}
\label{eq:rhos1}
\rho_\mathrm{sc}(T) = w \rho_\mathrm{sc}^{\Delta^s}(T) + (1-w) \rho_\mathrm{sc}^{\Delta^l}(T).
\end{equation}
As in the specific-heat case, $\rho_\mathrm{sc}^{\Delta^s}$ and 
$\rho_\mathrm{sc}^{\Delta^l}$ are the superfluid densities related to 
the small ($\Delta^s$) and large ($\Delta^l$) gaps, and $w$ is a 
relative weight. For each gap, $\rho_\mathrm{sc}(T)$ is given by: 
\begin{equation}
\label{eq:rhos}
\rho_\mathrm{sc}(T) =  1 + 2\int^{\infty}_{\Delta(T)} \frac{E}{\sqrt{E^2-\Delta^2(T)}} \frac{\partial f}{\partial E} \mathrm{d}E, 
\end{equation}
where $f$ and $\Delta$ are the Fermi- and the gap function, respectively, 
as in section~\ref{ssec:Cp_zero}.  
Here, the gap value at zero temperature $\Delta_0$ is the only adjustable parameter. 
As can be seen in figure~\ref{fig:lambda}(a), for TF-40\,mT, the 
temperature-independent behavior of $\lambda^{-2}(T)$ is consistent 
with an $s$-wave model with either a single- (solid line) or two gaps 
(dash-dotted line). The single-gap model, however, shows a less good 
agreement with the measured $\lambda^{-2}(T)$, as confirmed by the 
larger $\chi^2_{r}$ value (goodness of fit) compared to the two-gap model. Such conclusion 
is also supported by the low-$T$ specific-heat data shown in 
figure~\ref{fig:Cp2} and figure~\ref{fig:Cp3}(b) and the upper critical 
field in figure~\ref{fig:Hc2_determ}.
For the two-gap model, the zero-temperature magnetic penetration 
depth is $\lambda_\mathrm{0} = 87(1)$\,nm and the estimated gap values 
are $\Delta_0^s$ = 0.83(1)\,meV and $\Delta_0^l$ = 1.35(1)\,meV, with 
a weight $w = 0.1$. The latter are consistent with the gap values 
obtained from specific-heat data.
For the single-gap model, the estimated gap value is $\Delta_0 = 1.33$\,meV, 
with the same $\lambda_\mathrm{0}$ as in the two-gap case. 
In the TF-$\mu$SR with $\mu_0H_\mathrm{appl} = 120$\,mT [see figure~\ref{fig:lambda}(b)], 
the applied %magnetic 
field suppresses the smaller gap [see details in figure~\ref{fig:Cp3}(b) 
and figure~\ref{fig:Hc2_determ}]. Hence the $\lambda^{-2}(T)$ dependence 
is consistent with a single-gap $s$-wave model, leading to 
$\lambda_\mathrm{0} = 104(1)$\,nm and $\Delta_{0} = 1.10(1)$\,meV.

\subsection{\label{ssec:ZF_muSR}Zero-field $\mu$SR}
%
%==== figure =============================%
\begin{figure}[ht]
	\centering
	\includegraphics[width=0.8\textwidth,angle=0]{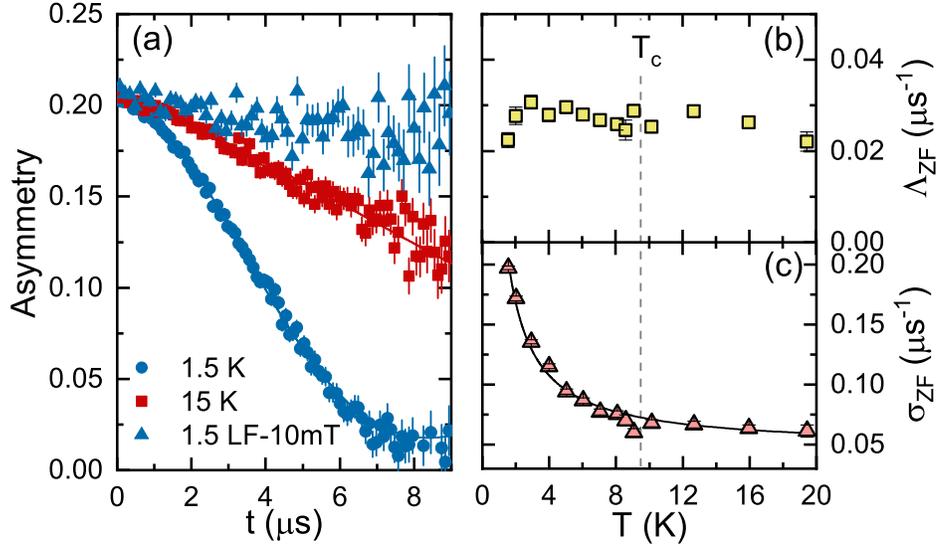}
	\vspace{-2ex}%
	\caption{\label{fig:ZF_muSR}(a) Representative ZF-$\mu$SR spectra for 
		ReBe$_{22}$ in the superconducting (1.5\,K) and the normal state (15\,K). 
		Additional LF-$\mu$SR data collected at 1.5\,K in a 10-mT applied field.  
		Solid lines are fits to equation (\ref{eq:KT_and_electr}). 
		Temperature dependence of the  
		Lorentzian- $\Lambda_\mathrm{ZF}$ (b), and Gaussian $\sigma_\mathrm{ZF}$ (c)
		relaxation rates. None of them shows clear 
		anomalies across $T_c$, marked by a dashed line. 
		The solid line in (c) represents a fit to 
		$\sigma_\mathrm{ZF}(T)  = 0.242\,T^{-1} -0.047$.}
\end{figure}
%=== end figure ==========================%

To search for a possible weak magnetism or TRS breaking in the superconducting 
state of ReBe$_{22}$, ZF-$\mu$SR measurements were performed in the 
1.5--20\,K temperature range. 
Normally, in the absence of external fields, there is no change in the ZF 
muon-spin relaxation rate near $T_c$. However, in case of a broken TRS, the 
onset of tiny spontaneous currents gives rise to associated (weak) magnetic fields, 
causing an increase in the muon-spin relaxation rate in the superconducting state. 
Representative ZF-$\mu$SR spectra for ReBe$_{22}$ collected above (15\,K) 
and below (1.5\,K) $T_c$ are shown in figure~\ref{fig:ZF_muSR}. 
No oscillations could be observed, implying a lack of magnetic order in 
ReBe$_{22}$. In such case, in absence of applied fields, the relaxation 
is mainly determined by the randomly oriented nuclear moments. 
Consequently, the ZF-$\mu$SR spectra of ReBe$_{22}$ can be modelled by 
means of a combined Lorentzian and Gaussian Kubo-Toyabe relaxation 
function~\cite{Kubo1967,Yaouanc2011}:
\begin{equation}
\label{eq:KT_and_electr}
A_\mathrm{ZF} = A_\mathrm{s}\left[\frac{1}{3} + \frac{2}{3}(1 -
\sigma_\mathrm{ZF}^{2}t^{2} - \Lambda_\mathrm{ZF} t)\,
\mathrm{e}^{\left(-\frac{\sigma_\mathrm{ZF}^{2}t^{2}}{2} - \Lambda_\mathrm{ZF} t\right)} \right] + A_\mathrm{bg}.
\end{equation}
Here $A_\mathrm{s}$ and $A_\mathrm{bg}$ are the same as in the 
TF-$\mu$SR case in equation (\ref{eq:TF_muSR}). In polycrystalline samples, the 1/3-nonrelaxing and 
2/3-relaxing components of the asymmetry correspond to the powder 
average of the local internal fields with respect to the initial 
muon-spin orientation. The resulting fit parameters vs.\ temperature, 
including the Lorentzian- $\Lambda_\mathrm{ZF}$
and Gaussian relaxation rates $\sigma_\mathrm{ZF}$, are shown in 
figures~\ref{fig:ZF_muSR}(b)-(c). Here $A_s$ was fixed to its average value 
of 0.205, however, the same features are also found in fits with released $A_s$.

The large relaxation rates reflect the significant 
nuclear magnetic moments present in ReBe$_{22}$. A similarly fast Gaussian 
relaxation was also found in other Re-based alloys~\cite{Shang2018,TianReNb2018}. 
This is in contrast to 
superconductors containing nuclei with small magnetic 
moments, as e.g., Mo$_3$Rh$_2$N~\cite{Shang2018Mo}, which exhibit a negligibly 
small relaxation.  
Despite the clear difference 
in the ZF-$\mu$SR spectra recorded in the normal and superconducting states %%
[figure~\ref{fig:ZF_muSR}(a)], neither $\Lambda_\mathrm{ZF}(T)$ nor 
$\sigma_\mathrm{ZF}(T)$ show distinct changes across $T_c$. The 
enhanced $\sigma_\mathrm{ZF}$ 
below 6\,K in figure~\ref{fig:ZF_muSR}(c) might be caused by tiny amounts of magnetic impurities, 
below the XRD and EDX detection threshold. This is also indicated 
by the Curie-Weiss-like behavior of $\sigma_\mathrm{ZF}(T)$ in figure~\ref{fig:ZF_muSR}(c), 
i.e., $\sigma_\mathrm{ZF} =  0.242 T^{-1} - 0.047$, 
whose positive curvature is opposite to the negative one, common in 
case of TRS breaking~\cite{Shang2018,TianReNb2018}. To further distinguish the intrinsic vs. extrinsic effects in $\sigma_\mathrm{ZF}(T)$,
 samples synthesized using even higher purity chemicals are desirable.
We also performed auxiliary longitudinal field (LF) $\mu$SR 
measurements at 1.5\,K. As shown in figure~\ref{fig:ZF_muSR}(a), a field of 10\,mT 
is already sufficient to lock the muon spins and to completely decouple them 
from the weak magnetic fields, confirming the sparse presence of magnetic impurities. 
In conclusion, the ZF-$\mu$SR results indicate a preserved TRS in the 
superconducting state of ReBe$_{22}$.

%==== figure =============================%
\begin{figure}[ht]
	\centering
	\includegraphics[width = 0.65\textwidth]{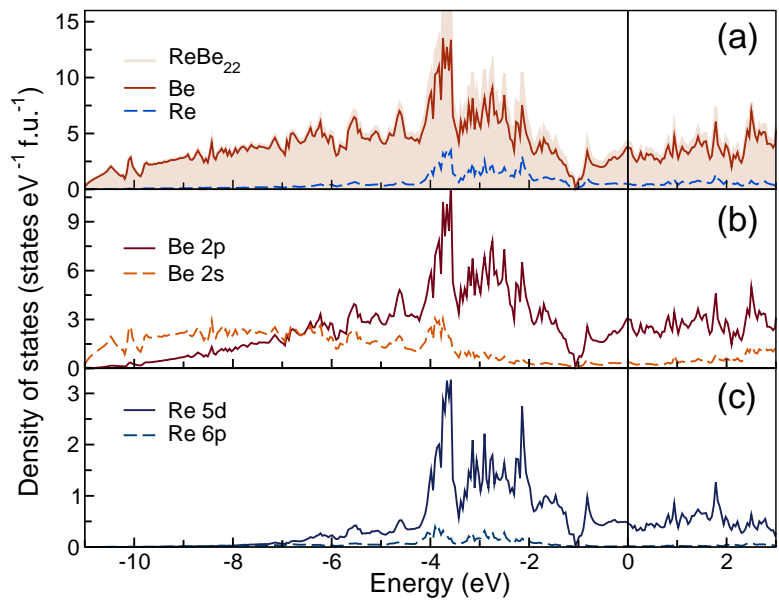}
	%\vspace{-2ex}%
	\caption{\label{fig:DOS} Calculated density of states 
		of ReBe$_{22}$ scaled to formula units (f.u.). 
		Total and partial (Be and Re) density of states (a). 
		Orbital-resolved density of states for the Be- (b) and Re atoms (c).}
\end{figure}
%== end figure ==========================%

\subsection{\label{ssec:DFT} Electronic band structure}

%
%==== figure =============================%
\begin{figure}[ht]
	\centering
	\includegraphics[width = 0.8\textwidth]{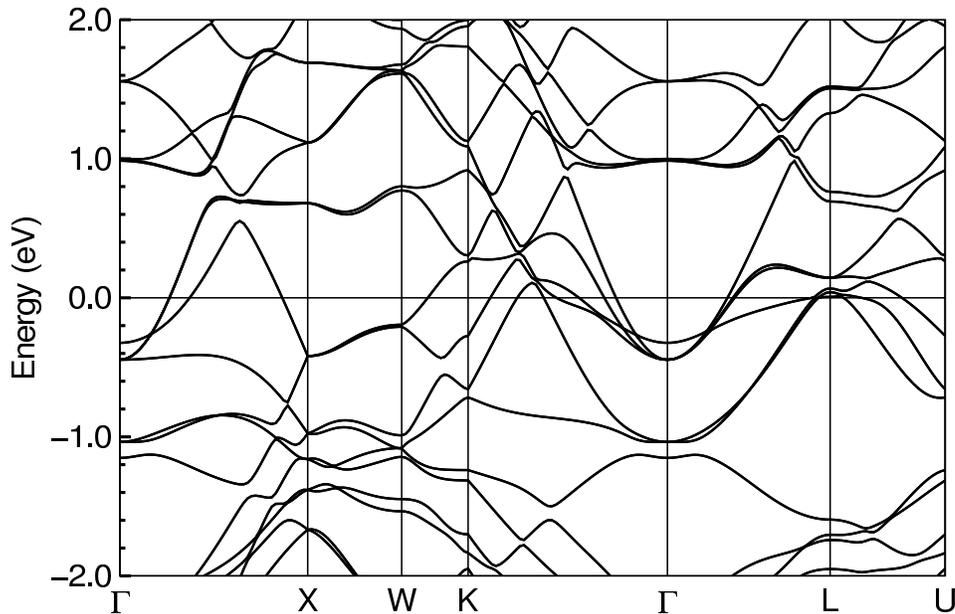}
	%\vspace{-2ex}%
	\caption{\label{fig:bands}Calculated electronic band structure of ReBe$_{22}$,  
		within $\pm$2\,eV from the Fermi energy level, neglecting 
		the spin-orbit coupling (here too small due to the low Re content).}
\end{figure}
%== end figure ==========================%
%

To shed more light on the underlying electronic properties of ReBe$_{22}$, 
we performed electronic band-structure calculations based on DFT, 
including spin-orbit coupling. Figure~\ref{fig:DOS} shows the total-, 
atomic-, and orbital-projected DOS, disclosing the metallic nature of 
the system through its nonzero DOS at the Fermi level. The main 
contributions to the latter arise from the Re-$d$ and Be-$p$ orbitals.  
While the Be-Be bonding is comprised primarily of 2$s$ orbitals, the 
Re-Be hybridization consists of Re-5$d$ and Be-2$p$ states. 
Notwithstanding a 4\%
Re-to-Be ratio in a ReBe$_{22}$ formula unit, Re atoms are over-represented with an almost 3 times larger 
weight of 12\% in the density of states at the Fermi level. 
Our calculations estimate a total DOS at the Fermi level 
of $N(\epsilon_\mathrm{F}) = 4$\, states/eV-f.u., comparable 
to the 3.25 states/eV-f.u.\ extracted from specific-heat data. 
Both values are significantly larger ($\sim$ 50 times) than that 
estimated for elemental Be~\cite{Bakai2007} and, consequently, may 
justify the huge increase in $T_c$ with respect 
to Be (from 0.026 to 9.4\,K). Interestingly, a similar $T_c$ value has 
been observed also when Be is deposited as a quenched condensed film~\cite{Granqvist1974}. 
Also in this case, the surge in $T_c$ was shown to originate from the 
increase of DOS at $E_\mathrm{F}$ in the structurally disordered 
condensate \cite{Bakai2007}. 

The ReBe$_{22}$ band structure shown in figure~\ref{fig:bands} reveals multiple dispersive 
bands crossing the Fermi energy. In particular, the electron pockets centered 
around the $\Gamma$ point are much larger than the hole pockets centered 
around the $L$ point. This circumstance is typical of multigap/multiband 
superconductors, as clearly reflected also in our experimental results.
Finally, the band splitting due to the spin-orbit coupling of Re is 
barely visible here due to the low Re/Be ratio.

%
%==== figure =============================%
\begin{figure}[ht]
	\centering
	\includegraphics[width = 0.7\textwidth]{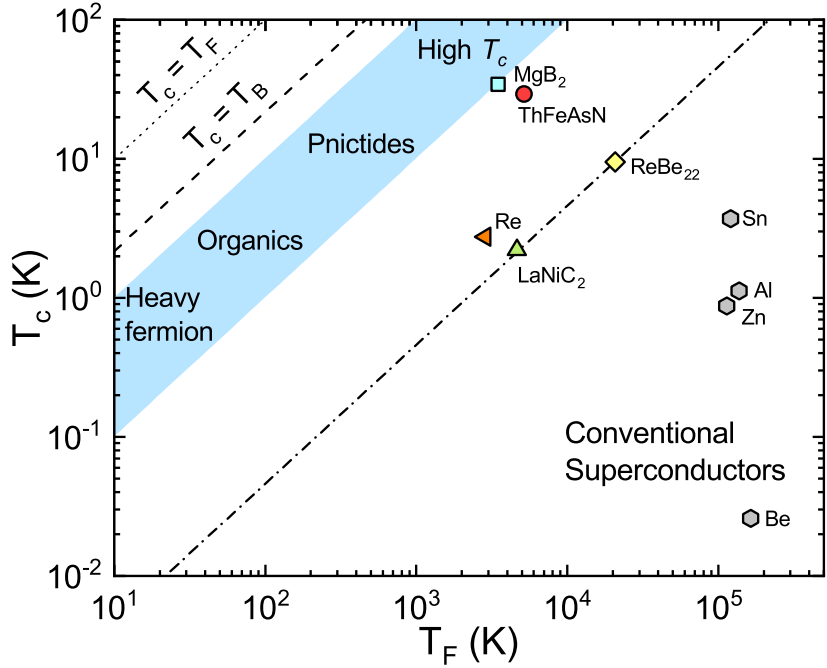}
	%\vspace{-2ex}%
	\caption{\label{fig:uemura}Uemura plot of the superconducting transition temperature 
		$T_c$ against the effective Fermi temperature $T_\mathrm{F}$ for different kinds of superconductors.   
		The shaded region, with  $1/100 < T_c/T_\mathrm{F}<1/10$,  indicates the band of unconventional 
		superconductors, such as heavy fermions,  organic, fullerenes, pnictides, cuprates, etc. 
		The dotted and dashed lines correspond to $T_c = T_\mathrm{F}$ and $T_c = T_\mathrm{B}$ 
		($T_\mathrm{B}$ is the Bose-Einstein condensation temperature), while the dash-dotted 
		line indicates $T_c/T_\mathrm{F} = 4.6 \times 10^{-4}$ for ReBe$_{22}$. 
		The data of the reference samples were adopted from references~\cite{Pietronero2018,Adroja2017,Shiroka2017,Barker2018,Uemura1991,TianReNb2018}.}
\end{figure}
%== end figure ==========================%

%==== Table =============================%
\begin{table}[!bht]
	\centering
	\caption{\label{tab:parameter} Normal-state and superconducting properties of ReBe$_{22}$, as 
		determined from electrical resistivity, magnetic susceptibility, specific-heat, and 
		$\mu$SR measurements. The London penetration depth $\lambda_\mathrm{L}$, 
		the effective mass $m^{\star}$, bare band-structure effective mass $m^{\star}_\mathrm{band}$, 
		carrier density $n_\mathrm{s}$, BCS coherence length $\xi_0$, electronic 
		mean-free path $l_e$, Fermi velocity $v_F$, and effective Fermi temperature
		$T_F$ were estimated following the equations~(40)--(50) in reference~\cite{Barker2018}.}  
		\begin{tabular}{lcp{12mm}lc}
		\mr
			Property                               & Value                    &     & Property                               & Value \\ \hline		
			$T_c^{\rho}$                           & 9.42(2)\,K               &   &  $\mu_0H_{c2}$                          & 0.60(3)\,T  \rule{0pt}{2.6ex} \\  
			$\rho_0$                               & 2.72(5)\,$\mu\Omega$cm   &    &  $\xi(0)$                               & 23(1)\,nm \\ 
			$\Theta_\mathrm{D}^\mathrm{R}$         &  590(5)\,K               &    & $T_c ^{\mu\mathrm{SR}}$(40\,mT)       & 8.8(1)\,K   \\ 
			$T_c^{\chi}$                           &  9.50(1)\,K              &    &  $w^{\mu\mathrm{SR}}$                   & 0.10   \\    
			$\mu_0H_{c1}$                          & 28.1(2)\,mT              &    &   $\Delta_0^s$$(\mu\mathrm{SR})$         & 0.83(1)\,meV    \\
			$\mu_0H_{c1}^{\mu\mathrm{SR}}$         & 24.9(5)\,mT              &    &  $\Delta_0^l(\mu\mathrm{SR})$           & 1.35(1)\,meV     \\
			$T_c^\mathrm{C}$                       & 9.36(2)\,K               &    &	 $\lambda_0$(40\,mT)                   &  87(1)\,nm      \\
			$\gamma_n$                             & 15.3(2)\,mJ/mol-K$^2$    &    & $\lambda_0$(120\,mT)                  & 104(1)\,nm   \\
			$\Theta_\mathrm{D}^\mathrm{C}$         & 545(5)\,K                &    &  $\lambda_\mathrm{L}$                   & 64(1)\,nm  \\	
			$\Theta_\mathrm{E}^\mathrm{C}$         & 792(5)\,K                &    & $\kappa$                               & 4.7(3)   \\
			$\lambda_\mathrm{ep}$                  & 0.64(1)                  &    & $m^{\star}$                            & 3.0(1)\,$m_e$        \\
			$N(\epsilon_\mathrm{F})$               & 3.25(3)\,eV-f.u.         &    &  $m_\mathrm{band}^{\star}$                      & 1.9(1)\,$m_e$   \\
			$N_\mathrm{band}(\epsilon_\mathrm{F})$ & 1.98(1)\,eV-f.u.         &    &   $\xi_0$                                & 53(1)\,nm    \\   
			$N(\epsilon_\mathrm{F})^\mathrm{DFT}$  & 4\,eV-f.u.               &    &   $l_e$                                  & 62(1)\,nm \\
			$w^C$                                  & 0.13                     &    &  $n_\mathrm{s}$                         & 2.06(5) $\times$ 10$^{28}$\,m$^{-3}$   \\
			$\Delta_0^s$$(C)$                      & 0.68(1)\,meV             &    &       $v_\mathrm{F}$                         & 3.29(7) $\times$ 10$^5$\,ms$^{-1}$   \\                             
			$\Delta_0^l(C)$                        & 1.43(1)\,meV             &    &     $T_\mathrm{F}$                         & 2.07(5) $\times$ 10$^4$\,K     \\                           
			$\Delta C/\gamma_\mathrm{n}T_c$        & 1.24(2)                  &    &                                             &\\   
\mr	
	\end{tabular}
\end{table}
%=== end table ==========================%

\subsection{\label{ssec:Dis} Discussion}

The different families of superconductors can be classified according to the ratio of 
the superconducting transition temperature $T_c$ to the effective Fermi temperature
$T_\mathrm{F}$, in a so-called Uemura plot~\cite{Uemura1991}. As can be seen in 
figure~\ref{fig:uemura}, several types of unconventional superconductors, including 
heavy-fermion, organic, high-$T_c$ iron pnictides, and cuprates, 
all lie in a $1/100 < T_c/T_\mathrm{F} <1/10$ range, here indicated by 
the shadowed region. Conversely, conventional BCS superconductors exhibit 
$T_c/T_\mathrm{F} <1/1000$, here exemplified by the elemental Sn, Al, and Zn.
Three typical examples of mul\-ti\-band superconductors, LaNiC$_2$, ThFeAsN, 
and MgB$_2$, are also shown in figure~\ref{fig:uemura}. According to the 
superconducting parameters obtained from our 
measurements (here summarized in table~\ref{tab:parameter}), the calculated $T_c/T_\mathrm{F}$ 
value for ReBe$_{22}$ is $9.5/20700 \sim 4.6 \times 10^{-4}$ 
(diamond in figure~\ref{fig:uemura}).
Although it cannot be classified as an unconventional superconductor, 
ReBe$_{22}$ is far away also from the region of conventional superconductors 
and shows practically the same ratio as the multiband superconductor 
LaNiC$_2$ (both lying in the same dash-dotted line). Compared to pure 
Be ($T_c/T_\mathrm{F} = 0.026/1.64 \times 10 ^5 = 1.58 \times 10^{-7}$)~\cite{Kittel2005,Falge1967}, 
the $T_c/T_\mathrm{F}$ value of ReBe$_{22}$ is enhanced due to the presence of 
diluted Re, the latter being characterized by a lower Fermi temperature and, hence, 
a larger $T_c/T_\mathrm{F}$ ratio (see Re in figure~\ref{fig:uemura}). 
Such conclusion is further supported by our electronic band-structure calculations, 
which show that, although Re contributes only 4\% to the atomic ratio, 
with its 12\% weight, it is over-represented in the density of 
states at the Fermi level. 

\vspace{7pt}
\section{\label{ssec:Sum} Conclusion}
To summarize, we investigated the physical properties of the ReBe$_{22}$ 
superconductor by means of electrical resistivity, magnetization, heat 
capacity, and $\mu$SR measurements, as well as  by electronic band-structure calculations.  
We find that ReBe$_{22}$ is a type-II superconductor ($\kappa \sim 4.7$), 
with a bulk $T_c \sim 9.4$\,K and critical fields $\mu_0H_{c1} = 28.1$\,mT 
and $\mu_0H_{c2} = 0.6$\,T.
The temperature dependence of the zero-field electronic specific heat 
and superfluid density reveal a nodeless superconductivity, well 
described by an isotropic $s$-wave model, which is more consistent with 
a multigap- rather than a single-gap superconductivity. 
The multigap features are further supported by the field-dependent 
electronic specific-heat coefficient, the upper critical field, and 
the calculated electronic band structure. The lack of spontaneous 
magnetic fields below $T_c$ indicates that, unlike in the Re-rich cases, 
in a Re-diluted superconductor such as ReBe$_{22}$, time-reversal symmetry 
is preserved. Compared to pure Be, the observed 400-fold increase in $T_c$ is shown 
to be due to the concomitant increase of both the electron-phonon coupling 
strength and of the density of states at the Fermi level. 
Future high-pressure studies of ReBe$_{22}$ should reveal the evolution 
of its superconducting properties upon decreasing the lattice parameter. 

\ack{The authors thank J.\ A.\ T.\ Verezhak for fruitful discussions and 
	the S$\mu$S beamline scientists for the assistance. 
	The DFT calculations were performed at the IFW-ITF cluster with the 
	assistance of U.\ Nitzsche.
	This work was supported by the National Key R\&D Program of China 
	(Grants No.\ 2016YFA0300202 and  No.\ 2017YFA0303100), National Natural 
	Science Foundation of China (Grants No.\ 11874320
	and No.\ U1632275), and  the Schwei\-ze\-rische Na\-ti\-o\-nal\-fonds 
	zur F\"{o}r\-de\-rung der Wis\-sen\-schaft\-lich\-en For\-schung, 
	SNF (Grants No.\ 200021-169455 and No.\ 206021-139082). }

\vspace{7pt}
\appendix
\section{EDX and field-dependent muon-spin relaxation}
%

%==== figure =============================%
\begin{figure}[th]
	\centering
	\includegraphics[width=0.7\textwidth,angle=0]{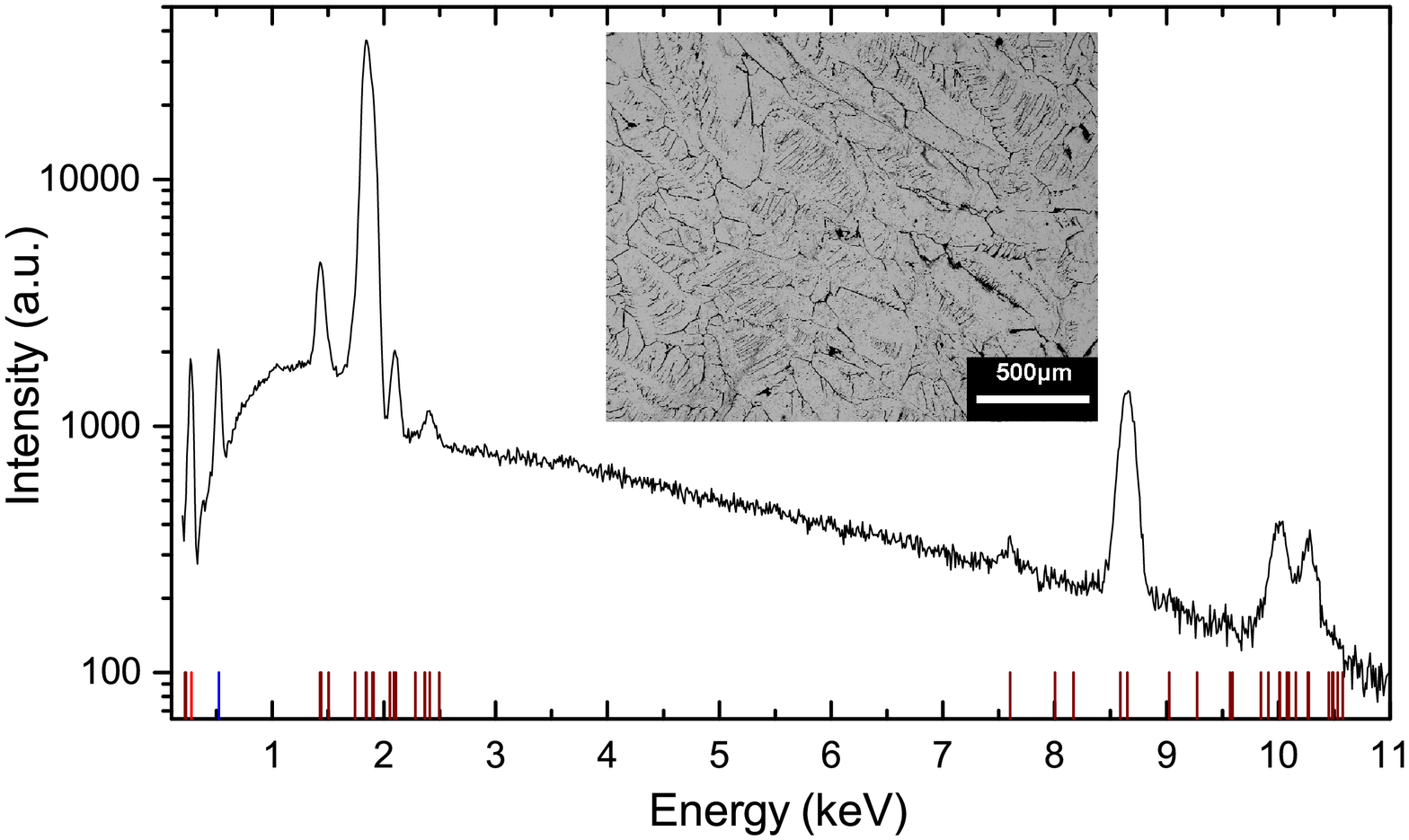}
	\vspace{-2ex}%
	\caption{\label{fig:edx}Energy-dispersive x-ray spectrum recorded on a ReBe$_{22}$ sample. 
		The x-ray emission lines attributed to rhenium, carbon, and oxygen are indicated by 
		maroon, red, and blue lines, respectively. Residual carbon and oxygen signals originate 
		from the atmosphere in the microscope chamber. 
		The absence of unidentified peaks reflects the high chemical purity of the sample. 
		The beryllium $K$-line is too low in energy (at 0.11\,keV) to be detected. 
		Note the logarithmic scale. Inset: electron micrograph (electron backscatter detector) of the investigated sample area. 
		The dark spots represent residual elemental beryllium inside the ReBe$_{22}$ grains.}
\end{figure}
%=== end figure ==========================%

%
\begin{figure}[!htp]
	\centering
	\includegraphics[width=0.8\textwidth,angle= 0]{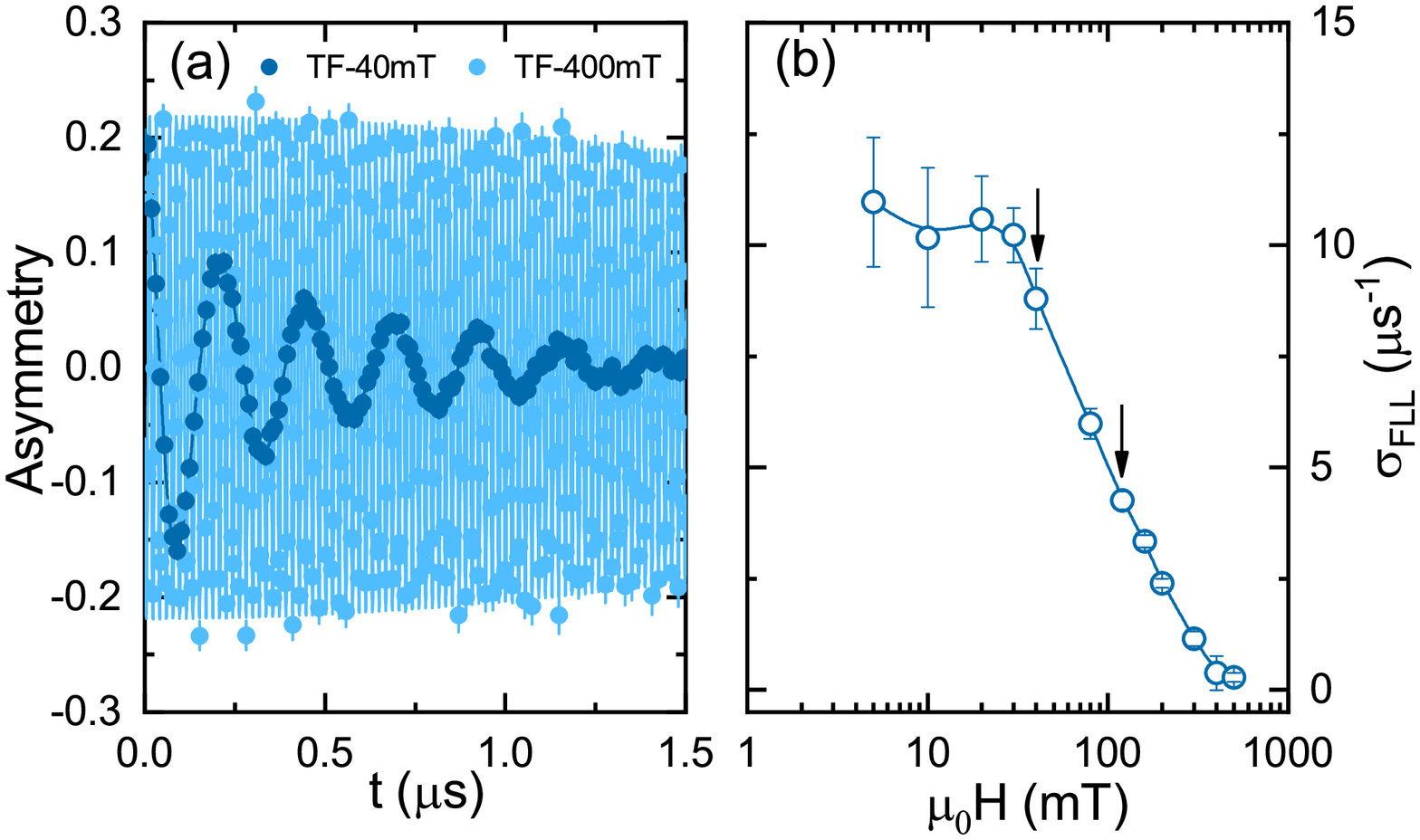}
	\caption{\label{fig:TF-muSR_H}(a) TF-$\mu$SR time spectra for 
		ReBe$_{22}$ measured at 1.5\,K (super\-con\-duct\-ing state) in 
		a field of 40 and 400\,mT. (b) Field-dependent superconducting Gaussian relaxation rate 
		$\sigma_\mathrm{FLL}(H)$. The arrows indicate the field values (40 
		and 120\,mT) chosen  for the temperature-dependent TF-$\mu$SR studies. 
		The solid line is a guide to the eyes.}
	%\label{fig:lines}
\end{figure}
%=== end figure ==========================%  
%

A typical energy-dispersive x-ray spectrum of ReBe$_{22}$ and the 
respective electron micrograph are shown in figure~\ref{fig:edx}. The 
high chemical purity of the sample is reflected in the lack of unknown peaks.

Figure~\ref{fig:TF-muSR_H}(a) shows the time-domain TF-$\mu$SR 
spectra of ReBe$_{22}$, collected in two applied fields, 40 and 400\,mT. 
The solid lines represent fits using the same model as that described in 
equation (\ref{eq:TF_muSR}). The resulting superconducting Gaussian 
relaxation rate $\sigma_\mathrm{FLL}(H)$ is summarized in 
figure~\ref{fig:TF-muSR_H}(b). Above the lower critical 
field $\mu_{0}H_{c1}$ (28.1\,mT), the Gaussian relaxation rate decreases 
continuously. By considering the decrease of the inter vortex distance 
with the field and the vortex-core effects, a field of 40\,mT was chosen 
for the temperature-dependent TF-$\mu$SR studies. For a comparison, 
the TF-$\mu$SR relaxation was also measured in a field of 120\,mT, 
here expected to suppress the small superconducting gap.

\section*{References}
\bibliography{ReBe22_bib}

\end{document}